\documentclass[10pt]{article}
\usepackage[utf8]{inputenc}
\pdfoutput=1
\let\OldS\S
\let\Oldc\c

\usepackage[utf8]{inputenc}
\usepackage{tikz, adjustbox, float} % For diagrams
\usepackage{amsmath, amsfonts, amssymb, amsthm, dsfont} % For math symbols
\usepackage{mathrsfs} % For additional caligraphic letters
\usepackage{soul} % For crossed text
\usepackage{bm} % For bold math so wha
\usepackage{booktabs}  % For better tables
\usepackage{enumitem} % For enumerate alphabet
\usepackage{mathtools}
\usepackage{mathabx}
\usepackage{physics}

\usepackage[linesnumbered,ruled,vlined]{algorithm2e}
\SetKwInOut{Input}{Input}
\SetKwInOut{Output}{Output}
\SetKwRepeat{Do}{do}{while}%

\usepackage{subcaption}
\usepackage{float}

\usepackage{titlesec}
\usepackage{cancel}
\titleformat{\paragraph}
{\normalfont\normalsize\scshape}{\theparagraph}{1em}{}
\titlespacing*{\paragraph}
{0pt}{3.25ex plus 1ex minus .2ex}{1.5ex plus .2ex}

\def\real{{\rm I\!R}}

\def\0{{\bf 0}}

\def\bt{{\bm{\theta}}}

\def\hsgm{\widehat\sigma}
\def\hsgm2{\hsgm^2}

\def\bOmega{{\bm{\Omega}}}

\def\X{{\bf X}}
\def\x{{\bf x}}

\DeclareMathOperator*{\argzero}{argzero}
\DeclareMathOperator*{\argsup}{argsup}

\usepackage{latexsym}
\usepackage{amsthm}
\usepackage{amsmath,amssymb,amsfonts,bm,amsbsy,bbm}
\usepackage{epsfig}
\usepackage{graphicx,rotating,lscape}
\usepackage[colorlinks=true,  urlcolor=blue, citecolor=blue, psdextra, pdfencoding=auto]{hyperref}
\usepackage{verbatim}
\usepackage{multirow,color}
\usepackage{geometry}
\usepackage{setspace}
\usepackage{bbm}

\usepackage{siunitx}

\def\X{{\bf X}}
\def\x{{\bf x}}

\def\S{{\bf S}}

\def\a{{\bf a}}
\def\b{{\bf b}}

\def\c{{\bf c}}

\def\bt{{\boldsymbol\theta}}

\def\btau{\boldsymbol\tau}

\def\0{{\bf 0}}

\def\bse{\begin{eqnarray*}}
	\def\ese{\end{eqnarray*}}
\def\be{\begin{eqnarray}}
	\def\ee{\end{eqnarray}}
\def\bsq{\begin{equation*}}
	\def\esq{\end{equation*}}
\def\bq{\begin{equation}}
	\def\eq{\end{equation}}

\def\boxit#1{\vbox{\hrule\hbox{\vrule\kern6pt  \vbox{\kern6pt#1\kern6pt}\kern6pt\vrule}\hrule}}
\def\bse{\begin{eqnarray*}}
	\def\ese{\end{eqnarray*}}
\def\be{\begin{eqnarray}}
	\def\ee{\end{eqnarray}}
\def\bsq{\begin{equation*}}
	\def\esq{\end{equation*}}
\def\bq{\begin{equation}}
	\def\eq{\end{equation}}

\def\a{{\bf a}}

\def\b{{\bf b}}

\def\X{{\bf X}}

\def\x{{\bf x}}

\def\log{\hbox{log}}

\def\squarebox#1{\hbox to #1{\hfill\vbox to #1{\vfill}}}
\def\btheta{{\boldsymbol \theta}}

\def\0{{\bf 0}}

\def\log{\hbox{log}}

\usepackage{bigints}

\newcommand*\wbar[1]{%
	\hbox{%
		\vbox{%
			\hrule height 0.5pt % The actual bar
			\kern0.3ex%         % Distance between bar and symbol
			\hbox{%
				\kern-0.1em%      % Shortening on the left side
				\ensuremath{#1}%
				\kern-0.1em%      % Shortening on the right side
			}%
		}%
	}%
}

\newtheoremstyle{mytheoremstyle} % name
{0.3cm}                      % Space above
{0cm}                        % Space below
{\itshape}                   % Body font
{}                           % Indent amount
{\bf}                   % Theorem head font
{: }                          % Punctuation after theorem head
{0em}                       % Space after theorem head
{}  % Theorem head spec (can be left empty, meaning normal)

\theoremstyle{mytheoremstyle}

\newtheorem*{Lemma*}{Lemma}

\newtheoremstyle{myExampleRemarkstyle} % name
{0.3cm}                    % Space above
{0cm}                           % Space below
{\itshape}                   % Body font
{}                           % Indent amount
{\bf}                   % Theorem head font
{: }                          % Punctuation after theorem head
{0em}                       % Space after theorem head
{}  % Theorem head spec (can be left empty, meaning normal)

\theoremstyle{myExampleRemarkstyle}

\newtheorem{innercustomgeneric}{\customgenericname}
\providecommand{\customgenericname}{}
\newcommand{\newcustomtheorem}[2]{%
	\newenvironment{#1}[1]
	{%
		\renewcommand\customgenericname{#2}%
		\renewcommand\theinnercustomgeneric{##1}%
		\innercustomgeneric
	}
	{\endinnercustomgeneric}
}

\newcustomtheorem{customthm}{Theorem}
\newcustomtheorem{customlemma}{Lemma}
\usepackage{mathalfa}
\usepackage{oubraces}

\newtheoremstyle{myExampleRemarkstyle} % name
{0.3cm}                    % Space above
{0cm}                           % Space below
{}                   % Body font
{}                           % Indent amount
{\bf}                   % Theorem head font
{: }                          % Punctuation after theorem head
{0em}                       % Space after theorem head
{}  % Theorem head spec (can be left empty, meaning normal)

\theoremstyle{myExampleRemarkstyle}

\let\refBKP\ref
\renewcommand{\ref}[1]{{\upshape\refBKP{#1}}}

\def\btheta{{\bm{\theta}}}
\def\htheta{\widehat{\theta}}
\def\hbtheta{\widehat{\bm{\theta}}}
\def\hbSigma{\widehat{\bm{\Sigma}}}
\def\hSigma{\widehat{\Sigma}}
\def\bSigma{\bm{\Sigma}}
\def\bOmega{\bm{\Omega}}
\def\hsigma{\widehat{\sigma}}

\def\btau{\bm{\tau}}

\def\blambda{\bm{\lambda}}
\def\bLambda{\bm{\Lambda}}
\def\bGamma{\bm{\Gamma}}

\usepackage{titlesec}
\titleformat{\section}{\normalfont\Large\scshape}{\thesection.}{1em}{}
\titleformat{\subsection}{\normalfont\large\scshape}{\thesubsection.}{1em}{}
\titleformat{\subsubsection}{\normalfont\scshape}{\thesubsubsection.}{1em}{}

\titlespacing*{\section}{0pt}{1em}{0em}
\titlespacing*{\subsection}{0pt}{1em}{0em}
\titlespacing*{\subsubsection}{0pt}{1em}{0em}

\usepackage[numbers,super,sort&compress]{natbib}
\usepackage[section]{placeins}

\usepackage{titlesec}

\titleformat*{\section}{\large \scshape}

\def\boxit#1{\vbox{\hrule\hbox{\vrule\kern2pt  \vbox{\kern2pt#1\kern2pt}\kern2pt\vrule}\hrule}}

\usepackage{tcolorbox}

\begin{document}

	%%%%% Title - section
	\begin{center}
		{\huge \textsc{Multivariate Adjustments for Average Equivalence Testing}}\\
		\vspace{0.5cm}
		
		$\textsc{Younes Boulaguiem}^{1,\star}$,  $\textsc{Luca Insolia}^{1,\star}$, \textsc{Maria-Pia Victoria-Feser}$^{2}$,\\ 
		\textsc{Dominique-Laurent Couturier}$^{3,4}$ \& \textsc{St\'ephane Guerrier}$^{1,5,6,\OldS}$ 
		\vspace{0.25cm}
	\end{center}
	
	%%%%%% Abstract + keyword section
	\begin{center}
		\footnotesize
		$^{1}$Geneva School of Economics and Management, University of Geneva, Switzerland;
		$^{2}$Department of Statistical Sciences, University of Bologna, Italy;
		$^{3}$Medical Research Council Biostatistics Unit, University of Cambridge, England;
		$^{4}$Cancer Research UK, Cambridge Institute, University of Cambridge, England;  
		$^{5}$School of Pharmaceutical Sciences, University of Geneva, Switzerland; \\
		$^{6}$Institute of Pharmaceutical Sciences of Western Switzerland, University of Geneva, Switzerland;\\
		$^{\star}$The first two authors contributed equally and are alphabetically ordered; $^{\OldS}$Corresponding author: \href{Stephane.Guerrier@unige.ch}{Stephane.Guerrier@unige.ch}.\\  
	\end{center} 
	
	\vspace{0.25cm}
	
	\textsc{Abstract:} 
	Multivariate (average) equivalence testing is widely used to assess whether the means of two conditions of interest are `equivalent' for different outcomes simultaneously. In pharmacological research for example, many regulatory agencies require the generic product and its brand-name counterpart to have equivalent means both for the AUC and C$_{\max}$ pharmacokinetics parameters. The multivariate Two One-Sided Tests (TOST) procedure is typically used in this context by checking if, outcome by outcome, the marginal $100(1-2\alpha$)\% confidence intervals for the difference in means between the two conditions of interest lies within pre-defined lower and upper equivalence limits. This procedure, already known to be conservative in the univariate case, leads to a rapid power loss when the number of outcomes increases, especially when one or more outcome variances are relatively large.
	In this work, we propose a finite-sample adjustment for this procedure, the multivariate $\alpha$-TOST, that consists in a correction of $\alpha$, the 
	significance level, taking the (arbitrary) dependence between the outcomes of interest into account and making it uniformly more powerful than the conventional multivariate TOST. We present an iterative algorithm allowing to efficiently define $\alpha^{\star}$, the corrected significance level, a task that proves challenging in the multivariate setting due to the inter-relationship between $\alpha^{\star}$ and the sets of values belonging to the null hypothesis space and defining the test size. We study the operating characteristics of the multivariate $\alpha$-TOST both theoretically and via an extensive simulation study considering cases relevant for real-world analyses --        i.e.,~relatively small
	sample sizes, unknown and possibly heterogeneous variances as well as 
	different correlation structures
	-- and show the superior finite-sample properties of the multivariate $\alpha$-TOST compared to its conventional counterpart. We finally re-visit a case study on ticlopidine hydrochloride and compare both methods when simultaneously assessing bioequivalence for multiple pharmacokinetic parameters.
	
	
	\par \noindent \textsc{Acknowledgements:}
	D.-L. Couturier is supported by UK Medical Research Council grant number MC\_UU\_00002/1 and by the Cancer Research UK grant C9545/A29580. L.~Insolia and S.~Guerrier are supported by the SNSF Grants \#176843 and \#211007 as well as by the Innosuisse Grants \#37308.1 IP-ENG and \#53622.1 IP-ENG. Y. Boulaguiem and M.-P. Victoria-Feser are partially supported by the SNSF Grant \#182684.
	
	\noindent \textsc{Conflict of Interest:} None declared.
	
	\noindent \textsc{Data availability statement:} 
	The data as well as an implementation of the method proposed in this article are available in the \texttt{cTOST} R package available on the GitHub repository: \href{https://github.com/stephaneguerrier/ctost}{\texttt{stephaneguerrier/cTOST}}.
	
	\section{Introduction}
	\label{sec:intro}
	
	Average equivalence tests, also known as similarity or parity tests, aim at assessing whether the means of two conditions or groups are equivalent for a given outcome. To do so, they require to define a tolerance region inside which the target parameter, typically the difference between the two means of interest, would lie to consider both groups to have similar means. This is different from traditional equality-of-means hypothesis tests in which the null and alternative hypotheses are inverted and in which the null hypothesis states that both means are equal rather than similar.  
	\par Historically, equivalence testing was studied by Lehmann\cite{Lehmann:1959} and Bondy,\cite{bondy1969test} and its applications to the field of pharmacokinetics \cite{westlake1972use,Metzler:74,Westlake:76} -- where it is often denoted as \textit{bioequivalence testing} --  boosted the interest in this area of research from the seventies. Nowadays, equivalence tests are widespread across several scientific and applied domains, such as social sciences,\cite{Lakens:17} imaging,\cite{SuXuErWaZh:22} medicine,\cite{OBrKi:22,WEHRLE2022,SaGiAuCpPa:22,BrEjXuWiHa:22} 
	economics,\cite{Feri:23} political sciences,\cite{aggarwal20232} 
	food sciences,\cite{MEYNERS2012} psychology,\cite{Lakens:18} sport sciences \cite{PhysiologyReview:22} and pharmaceutical sciences,\cite{WaLiChChSm:22} just to name a few. 
	\par In many situations, the interest lies in \textit{simultaneously} assessing whether the means of two conditions are ‘equivalent’ for different outcomes.\cite{linsinger1998influence,fogarty2014equivalence,knight2020forearm,hiraishi2021concordance,shieh2021improved,wable2021assessing,leday2022multivariate,aoki2023ultrastructural,gobel2023item} In pharmacological research for example,\cite{marzo2002bioequivalence,hoffelder2019equivalence,loingeville2020new} the formal approval of generic medicinal products is strongly facilitated if a sponsor is able to prove that the generic drug and its brand-name counterpart have equivalent means for different summary statistics of the relationship between drug concentration and time in bioavailability studies, like the area under the curve (AUC), the maximum concentration level (C$_{\max}$) and the time to reach C$_{\max}$ (t$_{\max}$).
	As these different pharmacokinetics parameters are estimated on the same samples, these quantities are typically correlated.
	\par In principle, multivariate equivalence testing procedures can be built upon a statistic summarising the similarity of the mean vectors of both conditions whilst taking the outcome dependence into account, such as the Mahalanobis distance.\cite{wellek2010testing} While such methods can provide a global measure of equivalence, they do not allow to assess equivalence for each individual outcome being tested.\cite{chervoneva2007multivariate} Indeed, in a case with the difference between the two mean vectors of interest equal to zero for all but one outcome, for which the difference would be substantial enough not to achieve marginal equivalence, a Mahalanobis distance-based overall equivalence assessment could still lead to the declaration of equivalence between the two mean vectors, as the difference in means measured on this single outcome would be diluted across dimensions. 
	Therefore, such approaches are not directly comparable and cannot be used in standard equivalence testing problems.
	\par For this reason, the Two One-Sided Tests (TOST) procedure,\cite{schuirmann1987comparison} traditionally used in the univariate case, is often also considered in the multivariate case, where it will be referred to as the (conventional) multivariate TOST. In the univariate case, equivalence is declared if the target parameter is both significantly greater than the lower, and significantly lower than the upper equivalence bounds at a given significance level $\alpha$. The most common way of assessing equivalence is then to use the {\it interval-inclusion principle} and check whether the $100(1 - 2\alpha$)\% confidence interval for the target parameter falls within these tolerance margins.\cite{wellek2010testing} 
	In the multivariate case, this intuitive and visual procedure is used for each outcome independently and equivalence is declared if the confidence interval of each outcome-related difference in means lies within the same pre-defined equivalence bounds as in the univariate case. 
	\par Pallmann and Jaki\cite{pallmann2017simultaneous} consider a multivariate assessment of equivalence based on simultaneous confidence sets for the target parameter vector, defined by means of different methods allowing to take the outcome dependence into account, like the non-parametric bootstrap, the lima\Oldc{c}on of Pascal,\citep{brown1995optimal} and Tseng's method.\citep{tseng2002optimal} The authors conclude that ``\textit{there is no confidence region that has consistently good power across all simulated scenarios}'' [p.~4596] and that ``\textit{the use of} [simultaneous confidence sets] \textit{will hardly ever bring about a power gain as compared to the conventional} [multivariate] \textit{TOST}'' [p.~4600]. We therefore focus on the multivariate TOST hereafter. 
	\par In the univariate case, the TOST's size (i.e., the type-I error rate) is known to be lower than the pre-specified significance level $\alpha$,\cite{berger1996bioequivalence} leading to a substantial loss in power, i.e., to a lower probability of declaring equivalence when this is true, so that several adjustments to this method have been proposed.\cite{brown1997unbiased} Recently, Boulaguiem et al.\cite{boulaguiem23} put forward the $\alpha$-TOST, an adjustment for the univariate TOST that guarantees the resulting test to be size-$\alpha$ and to have a higher power than the conventional TOST. In the multivariate case, Pallmann and Jaki\cite{pallmann2017simultaneous} show that the test size of the conventional multivariate TOST sharply decreases when the number of outcomes increases even when outcomes are strongly correlated, meaning that its loss in power is considerably larger in the multivariate setting. Furthermore, like its univariate counterpart, its power is also strongly affected by relatively large outcome variances, with power levels often reaching zero, hence making it impossible to declare equivalence when it is actually true. 
	\par 
	Therefore, in this work, we extend the (univariate) $\alpha$-TOST procedure of Boulaguiem et al.\cite{boulaguiem23} to the multivariate case and introduce the multivariate $\alpha$-TOST. This method relies on a finite-sample adjustment of the multivariate TOST allowing to take the arbitrary outcome dependence into account and leading to a better test size and increased power, making the multivariate $\alpha$-TOST uniformly more powerful than its conventional counterpart. We show that our finite-sample adjustment is exactly size $\alpha$ when performed at the population level. 
	Moreover, the proposed method has an empirical size (i.e.,~when considering a sample version of the adjustment) differing from $\alpha$ by a higher-order term that is asymptotically negligible in the sense that this term converges considerably faster than the other random variables involved in the procedure.
	This is confirmed by our extensive simulation study, showing that the multivariate $\alpha$-TOST achieves an empirical size that is closer to the nominal level $\alpha$ compared to the multivariate TOST, thus leading to an increased power.
	This extension of the $\alpha$-TOST to the multivariate setting is challenging due to the difficulty in correctly specifying the test size, which relates to finding the set of target parameter values maximising the probability of declaring equivalence under the null hypothesis. In general, this set cannot be characterised in closed form, making the problem considerably more involved than the univariate $\alpha$-TOST.
	Also, as this set of values is a function of $\alpha$, a recursive algorithm was needed to obtain $\alpha^{\star}$, the corrected significance level. 
	\par The paper is organised as follows. The multivariate equivalence framework is introduced in Section~\ref{sec:background}. The multivariate TOST procedure and its operating characteristics are described in Section~\ref{sec:conventionaltost}. Our proposed multivariate $\alpha$-TOST, its theoretical properties, as well as the algorithm we developed to compute the adjusted significance level are presented in Section~\ref{sec:proposal}. The results of the extensive simulation study we performed to compare the performance of both methods are discussed in Section~\ref{sec:simulation}. Section~\ref{sec:application} re-visits the ticlopidine hydrochloride case study \cite{marzo2002bioequivalence, pallmann2017simultaneous} and compares the results of both methods when simultaneously assessing bioequivalence for multiple pharmacokinetic parameters. Finally,  potential extensions and directions for future research are discussed in Section~\ref{sec:conclusion}.

	\section{Multivariate Equivalence} 
	\label{sec:background}
	
	Throughout the paper, we denote by bold lower and upper case letters respectively vectors and matrices. Moreover, we use $x_i$, $X_{i,j}$, and $X_{i}$ to refer to the $i$-th element of the vector $\x$, the $(i,j)$-th element of the matrix $\X$ and the $i$-th diagonal element of $\X$, respectively.
	
	Let $\btheta$ denote the $m$-dimensional vector of unknown target population parameters, such as the difference in means between two experimental conditions, for each of the $m$ outcomes of interest, and let $\hbtheta=[\htheta_1, \ldots, \htheta_m]^T$ be an unbiased estimator of $\btheta$. Then, let $\bm{\Sigma}$ denote the unknown ($m \times m$) covariance matrix of $\hbtheta$, which depends on the number of degrees of freedom $\nu$ and that we suppose to be positive definite. Finally let $\hbSigma$ be an unbiased estimator of  $\bSigma$, which we also suppose to be positive definite for simplicity.
	
	We consider the following canonical form of the multivariate equivalence problem\cite{wang1999statistical}: 
	\begin{equation} \label{eq:canon}
		\hbtheta \sim \mathcal{N}_m \left( \btheta, \bm{\Sigma} \right)  
		\quad \text{and} \quad 
		\nu \hbSigma \sim \mathcal{W}_m \left(\nu, \bm{\Sigma} \right)   ,
	\end{equation}
	where $\hbtheta$ and $\hbSigma$ are independent, and $\mathcal{N}_m(\cdot)$ and $\mathcal{W}_m(\cdot)$ respectively represent the $m$-variate normal and Wishart distributions. Furthermore, we define $ \sigma_{j}^2 \vcentcolon= \Sigma_{j}$, the variance of $\htheta_j$. The canonical form presented in (\ref{eq:canon}) is design-agnostic as it includes parallel or (replicated) crossover designs, for example.

	The hypotheses of interest to assess equivalence for each $\theta_j$ parameter, where $j=1,\ldots,m$, are given by
	\begin{equation} \label{eq:hypotheses_mult}
		\begin{alignedat}{2}
			&\text{H}_{0}: \theta_j \leq a_j \quad 
			\text{or} & \quad
			& \theta_j \geq b_j \quad 
			\text{ for some $j$,
			} \\ 
			&\text{H}_{1}: \theta_j > a_j \quad 
			\text{and} & \quad
			& \theta_j < b_j  \quad 
			\text{ for all $j$, 
			} 
		\end{alignedat}
	\end{equation}
	where $\a \vcentcolon= [ a_1, \ldots,  a_m ]^T$ and $\b \vcentcolon= [ b_{1}, \ldots, b_{m} ]^T $ respectively denote the (marginal) lower and upper equivalence margins. These margins are considered as known $m$-variate constants typically given by the context of the analysis. For example, regulatory agencies often rely on $a_j=\log(0.8)$ and $b_j=\log(1.25)$ to assess bioequivalence of medicinal products.\cite{EnTo:19}
	The null and alternative hypotheses in \eqref{eq:hypotheses_mult} can be expressed in a multivariate manner as follows
	\begin{equation}  \label{eq:hypotheses_mult2}
		\text{H}_0: \; \btheta \not\in \bm{\Theta}_1 \quad \text{vs.} \quad \text{H}_1: \; \btheta \in \bm{\Theta}_1,
	\end{equation}
	where $ \bm{\Theta}_1 \vcentcolon= \{\x\in\real^m\,\big|\,a_j < x_j < b_j,\, j=1,\dots,m\} $ defines the $m$-dimensional parallelotope delimited by the equivalence margins. 
	For simplicity,
	in the following we assume that 
	equivalence margins are symmetric around zero and have the same value for all $j$ with
	$
	\c \vcentcolon= [c, \ldots, c ]^T = \b = -\a $, so that
	$\bm{\Theta}_1 = \{ \x \in\real^m\,\big|\,| x_j| < c,\, j=1,\dots,m\}$.
	
	\section{Multivariate TOST}
	\label{sec:conventionaltost}
	
	Multivariate equivalence is typically assessed through the TOST formulation,\cite{schuirmann1987comparison} where equivalence is declared when all outcome-related 100$(1- 2 \alpha)$\% confidence intervals for the difference in means are contained within the $m$-dimensional hypercube defined by $\bm{\Theta}_1$. 
	Importantly, Berger\cite{berger1982interunion} proved that multivariate equivalence at level $\alpha$ can be assessed through multiple univariate tests. 
	Namely, to test for equivalence in \eqref{eq:hypotheses_mult2}, the multivariate TOST is based on the test statistics
	$$
	T_{l_{j}} \vcentcolon= \frac{\htheta_j + c}{\widehat{\sigma}_{j}} 
	\quad \text{and} \quad 
	T_{u_j} \vcentcolon= \frac{\htheta_j - c}{\widehat{\sigma}_{j}}, 
	$$
	for $ j=1,\ldots,m$.
	Hence, at a significance level $\alpha$, multivariate equivalence is declared if each $j$-th marginal test statistic for $j = 1, \ldots, m $ satisfies
	\begin{equation}
		\label{eqn:tstat-assess}
		T_{l_j} \geq t_{\alpha, \nu} 
		\quad \text{and} \quad 
		T_{u_j} \leq -t_{\alpha, \nu},
	\end{equation}
	where
	$t_{\alpha, \nu}$ represents the upper $\alpha$ quantile of a Student's $t$-distribution with $\nu$ degrees of freedom.
	
	Using the \textit{interval-inclusion principle},\cite{wellek2010testing} the rejection region associated with the multivariate TOST, for any $\hbSigma$, is given by
	\begin{equation} \label{eq:TOST_rej_region}
		C_1(\hbSigma) \vcentcolon= \bigcap_{j=1}^m \left\{ \lvert \htheta_j \rvert \leq c - t_{\alpha,\nu}  \hsigma_{j} \right\} ,
	\end{equation}
	showing that equivalence can never be declared if 
	\begin{equation} \label{eq:TOST_sigma_ineq}
		\widehat{\sigma}_{j} >
		M_j
		\vcentcolon= c/t_{\alpha,\nu},
	\end{equation}
	for any $j$, even though $\htheta_j=0$. This result has important implications on the test size of the multivariate TOST that we discuss in Section~\ref{sec:MTost-power}, after discussing the probability of declaring equivalence which for simplicity we refer to as the power function of this procedure.
	
	\subsection{Power function of the Multivariate TOST}
	
	In the univariate case, Phillips\cite{Phil:90} showed that the probability of declaring equivalence that arises from  \eqref{eq:TOST_rej_region} is related to the bivariate non-central Student's $t$-distribution and can be expressed as a special case of Owen's Q-functions.\cite{Owen:65} To the best of our knowledge, this link is not available in multivariate settings (i.e.,~$m>1$). To define the power of the multivariate TOST, we therefore consider a general approach relying on the integration of the joint density of $ \hbtheta$ and $\widehat{\bm{\Sigma}}$ over the 
	region 
	of interest, which can be obtained using \eqref{eq:TOST_rej_region}-\eqref{eq:TOST_sigma_ineq} taking into account that $\widehat{\bm{\Sigma}}$ is a positive definite matrix. Let
	$$
	\Delta(\widehat{\bm{\Sigma}} \lvert 
	\bm{\Sigma}, \bt) \vcentcolon=  
	\int_{k^l_m}^{k^u_m} \cdots \int_{k^l_1}^{k^u_1} 
	f (\hbtheta \lvert \btheta, \bSigma ) ~ d\htheta_1 \ldots d \htheta_m
	$$
	denote the probability associated to the $m$-dimensional normal distribution centred at $\btheta$ with covariance $\bSigma$, whose density is denoted by $f(\cdot|\btheta,\bSigma)$, over the limits of integration 
	$k_j^l \vcentcolon= -c + t_{\alpha, \nu}  \widehat{\sigma}_{j}$, and
	$k_j^u \vcentcolon= c - t_{\alpha, \nu} \widehat{\sigma}_{j}$, and let $g(\cdot \lvert \bSigma )$ denote the density of an $m$-dimensional Wishart distribution with scale matrix $\bSigma$. Then, the probability of rejecting H$_0$ for
	the multivariate TOST can be expressed as\cite{phillips2009power}:
	\begin{equation} 
		\begin{aligned} \label{eq:TOST_power}
			p(& \alpha,  \btheta, \bSigma, \nu , \bm{c} ) \vcentcolon=
			\Pr \left\{  
			C_1(\hbSigma) 
			\lvert 
			\alpha, \btheta, \bSigma, \nu , \bm{c}
			\right\}     \\
			&= 
			\int_0^{M_m^2} \cdots \int_0^{M_1^2} 
			\int_{-M_{m-1} M_m}^{M_{m-1} M_m } \cdots \int_{-M_1 M_2}^{M_1 M_2} 
			\eta( \widehat{\bm{\Sigma}} ) 
			\Delta(\widehat{\bm{\Sigma}} \lvert \bm{\Sigma}, \bt) 
			g ( \hbSigma \lvert \bSigma )
			~ d\hSigma_{1,2} \ldots d\hSigma_{m-1,m} 
			d\hsigma_{1}^2 \ldots d\hsigma_{m}^2 ,
		\end{aligned}
	\end{equation}
	for 
	$$
	\eta( \hbSigma ) \vcentcolon= 
	I( \lvert \hSigma_{1,2} \rvert < \hsigma_{1} \hsigma_{2} ) 
	\times \cdots \times
	I( \lvert \hSigma_{m-1,m} \rvert < \hsigma_{m-1} \hsigma_{m} ) ,
	$$
	where $C_1(\hbSigma)$ is given in \eqref{eq:TOST_rej_region}, $M_j$ in \eqref{eq:TOST_sigma_ineq}, and 
	$I(\cdot)$ denotes the indicator function. The power function in \eqref{eq:TOST_power} can be computed via Monte Carlo integration, for example, leading to a more computationally intensive procedure than in the univariate case, but without substantially impacting the computational burden as the Owen's Q-functions used in the univariate case are typically solved by numerical integration. 
	
	In the next section, we show how the multivariate TOST becomes increasingly conservative as $m$ increases, hence justifying our finite-sample correction.

	\subsection{Test size of the Multivariate TOST}
	\label{sec:MTost-power}
	
	The multivariate TOST’s size is a function of $\alpha, \btheta, \bSigma$, $\nu $ and $\bm{c}$, and is defined as the supremum of \eqref{eq:TOST_power} over the space of the null hypothesis.\cite{Lehmann:1986} 
	We define
	\begin{equation} \label{eq:lambda}
		\blambda = \left[ \lambda_1, \ldots, \lambda_m \right]^T \in \bLambda(\alpha, \bSigma) \vcentcolon= \argsup_{\btheta \notin \bm{\Theta}_1 } \;
		p(\alpha, \btheta, \bSigma, \nu , \bm{c} ) .
	\end{equation}
	The vector $\blambda$ depends on both $\alpha$ and $\bSigma$ as highlighted by the definition of $\bLambda(\alpha, \bSigma)$.
	To simplify the notation, these dependencies are generally omitted except when needed to avoid ambiguity.
	Therefore, the size is given by
	\begin{align} \label{eq:size}
		p ( \alpha, \blambda, \bSigma, \nu , \bm{c} ) 
		= \sup_{\btheta \notin \bm{\Theta}_1 } 
		p(\alpha, \btheta, \bSigma, \nu , \bm{c} ) .
	\end{align}

\begin{figure}
		\centering \includegraphics[width=1\linewidth]{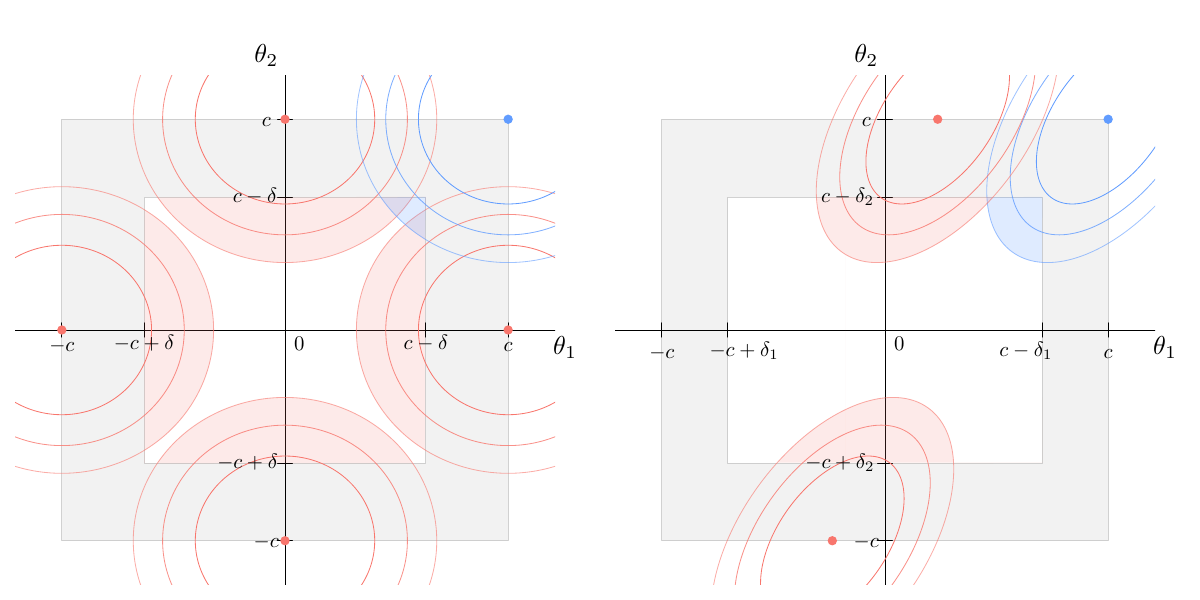}
		\caption{Target parameter space in two bivariate equivalence scenarios, respectively considering independent and homoscedastic (left panel) and dependent and heteroscedastic (right panel) $\htheta_j$. The outer square margins, taking the values $-c$ or $c$, correspond to the boundary between the spaces of the null and alternative hypotheses. The inner (white) rectangle, with limits depending on $c$ and $\delta=t_{\alpha,\nu} \widehat{\sigma}$ in the homoscedastic case, and $\delta_j=t_{\alpha,\nu} \widehat{\sigma}_j$ with $j=1,2$ in the heteroscedastic case, corresponds to the region in which 
			$\hbtheta$ should lie for equivalence to be accepted.
			The red dots correspond to the coordinates of the difference in means (y- and x- axes) leading to the supremum of the probability of declaring equivalence over the space of the null hypothesis, to which the red-shaded areas in the inner white rectangle provide a visual representation. In each panel, the blue dot corresponds to the $(c,c)$ coordinates sometimes used in the literature to study the multivariate TOST test size and that does not correspond to the supremum of \eqref{eq:size} except when $\rho\to1$, where $\rho$ denotes the correlation, and variances are equal.}
		\label{fig:argsups}
	\end{figure}

	The set $\bLambda(\alpha, \bSigma)$ contains multiple solutions and the supremum of the probability of declaring equivalence over the space of the null depends on the level of the test and the covariance matrix, making the test size of the multivariate TOST not straightforward to compute. Figure \ref{fig:argsups} showcases this difficulty with two bivariate cases considering independent and homoscedastic (left panel) and dependent and heteroscedastic (right panel) $\htheta_j$ elements. In both 
	panels, the red dots correspond to the set of values of the two differences in means (y- and x- axes) leading to the supremum of the probability of declaring equivalence over the space of the null, where the red-shaded areas in the inner white square provide a visual representation of this probability. In the independent and homoscedastic case, $\bLambda(\alpha, \bSigma)$ has four solutions corresponding to coordinates where either component is equal to 0  while the other one is equal to $-c$ or $c$. In the dependent and heteroscedastic case shown here, $\bLambda(\alpha, \bSigma)$ has two solutions corresponding to coordinates where the component associated to the estimate with the largest variance equals $-c$ or $c$ while the other one is slightly smaller or greater than $0$, respectively. This contradicts the use of the $(c,c)$ coordinate (blue point) to define the multivariate TOST size, as it corresponds to the supremum of (\ref{eq:size}) only in very specific settings, namely when elements of $\hbtheta$ are homoscedastic and have a correlation $\rho$ approaching one.

	From \eqref{eq:size}, we can deduce that the multivariate TOST is conservative by noting that 
	\begin{equation}
		\begin{aligned}
			p ( \alpha, \blambda, \bSigma, \nu , \bm{c} ) &= \Pr \left(  
			\bigcap_{j=1}^m \left\{ \lvert \htheta_j \rvert < c - t_{\alpha,\nu}  \hsigma_{j} \right\}
			\right) 
			\leq \min_{j=1, \ldots, m} \, \Pr \left(  
			\left\{ \lvert \htheta_j \rvert < c - t_{\alpha,\nu}  \hsigma_{j} \right\}
			\right)\\
			&\leq \Pr \left(  
			\left\{ \lvert \htheta_h \rvert < c - t_{\alpha,\nu}  \hsigma_{h} \right\}
			\right) 
			<
			\alpha.
		\end{aligned}
		\label{eq:level:tost}
	\end{equation}
	The first inequality corresponds to one of Fréchet inequalities, the second inequality is a consequence of $\blambda \notin \bm{\Theta}_1$, and thus, at least one of the entries of this vector, whose index is denoted as $h$, is either equal to $c$ or $-c$. Finally, the last inequality is a property of the univariate TOST.\cite{Deng2020}
	
\begin{figure}
		\centering
		\includegraphics[scale=0.8]{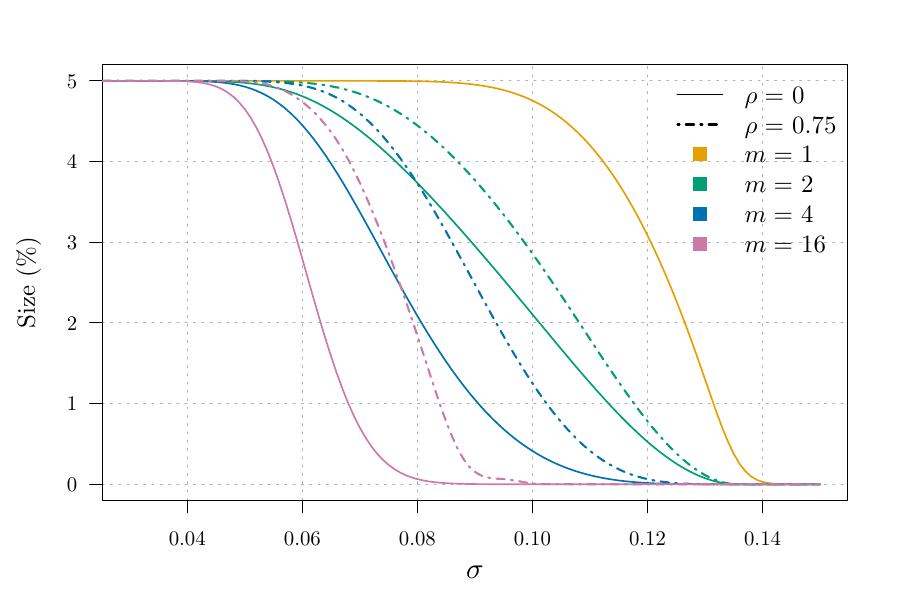}
		\caption{Size of the multivariate TOST (y-axis) as a function of $\sigma$ (x-axis) for different dimensions $m$ (coloured lines) and correlations $\rho$ (line type) under homoscedasticity and known variance, with $c = \log(1.25)$ and $\alpha = 0.05$.}  
		\label{fig:size:sigma:rho}
	\end{figure}

	If the conventional multivariate TOST is size-$\alpha$ asymptotically (refer to Appendix~\ref{appendix:size_TOST} for details), the conservative nature of the TOST in finite sample, well known in the univariate case,\cite{Deng2020} is exacerbated in multivariate settings. To illustrate this, we consider a homoscedastic setting (i.e.,~$\sigma \vcentcolon= \sigma_1 = \ldots = \sigma_m$) under two forms of dependence (equi-correlated with correlation $\rho \in \{0, 0.75\}$) when $\bSigma$ is known, for simplicity. Similar results, not reported here for brevity, hold for the case when $\bSigma$ is estimated.
	Figure~\ref{fig:size:sigma:rho} shows the size of the multivariate TOST as a function of $\sigma$ for increasing number of outcomes $m$ and two correlation levels.
	We can note that, as $\sigma$ increases, the test size sharply decreases especially when the number of outcomes is large, and that this decrease is even stronger under independence, leading to a loss of power of this procedure. 
	Under independence and when $\bSigma$ is known, we can derive the size function analytically. In this case, $\bSigma = \sigma^2 \mathbf{I}_m$ with $\sigma^2$ known. We have
	$$
	\blambda = [ c, 0, \ldots, 0 ] ^T \in \bLambda(\alpha,  \sigma^2 \mathbf{I}_m) = \{ \x \in \real^m \,\big|\,  \lVert \x \rVert_1 = c, ~\lVert \x \rVert_0 =  1  \} ,
	$$ 
	where $\lVert \cdot \rVert_1$ is the $L_1$-norm and $\lVert \cdot \rVert_0$ is the pseudo $L_0$-norm. 
	Then, the test size is
	\begin{equation}
		\label{eq:size_multiv_tost}
		\begin{aligned}
			p \{ \alpha, \blambda, \sigma^2\mathbf{I}_m , \bm{c} \} =
			\left\{1 - \Phi\left( z_{\alpha}\right) - \Phi\left( z_{\alpha}  - \frac{2 c }{{\sigma}}\right)\right\}
			\left\{1 - 2\Phi\left( z_{\alpha}  - \frac{c}{{\sigma}}\right)\right\}^{m-1}  ,
		\end{aligned}
	\end{equation}
	where $z_{\alpha}$ denotes the upper $\alpha$ quantile of a standard normal distribution and $\Phi$ denotes its cumulative distribution function. Whilst the first term on the right-hand side of \eqref{eq:size_multiv_tost} represents the size of the univariate TOST with known variance $\sigma^2$, the second term accounts for the multivariate nature of the problem and leads to a decreased test size as $m$ increases  (see Appendix~\ref{appendix:simple_TOST} for details on the derivation of \eqref{eq:size_multiv_tost}).

	\section{The Multivariate $\alpha$-TOST}
	\label{sec:proposal}
	
	In this section, we propose an adjustment of the multivariate TOST that leads to a better test size and therefore to a higher power. Our approach consists in a generalisation of the univariate $\alpha$-TOST to the $m$-dimensional framework ($m\geq1)$ and operates by shrinking the marginal confidence intervals for each outcome-related target parameter by adjusting the level of the test in a way that guarantees the resulting procedure to be size-$\alpha$
	when considering an adjustment based on $\bSigma$.
	Moreover, when this adjustment is based on $ \hbSigma$ instead of $\bSigma$, we show that the size differs from $\alpha$ by a higher-order term that is negligible compared to other sources of variability involved in the procedure.

	\subsection{Multivariate Test Level Adjustment}
	\label{sec:MalphaTOST}
	
	To obtain an alternative equivalence test that is size $\alpha$, we define $\alpha^*$, the adjusted level used to assess the hypotheses in \eqref{eq:hypotheses_mult2}, as
	\begin{equation} 
		\label{eq:alpha_TOST_pop}
		\alpha^*(\bOmega) \vcentcolon=\argzero_{\gamma \in [\alpha, 0.5)} \;\big[     p\{\gamma, \blambda(\gamma, \bOmega), \bOmega, \nu , \bm{c} \} - \alpha\big],
	\end{equation}
	where $\blambda(\gamma, \bOmega)$ is given in \eqref{eq:lambda}, and $\bOmega$ indicates a general covariance matrix which could correspond to the theoretical one, $\bm{\Sigma}$, or its empirical counterpart, $\widehat{\bm{\Sigma}}$. For simplicity,  we respectively define $ \alpha^* \vcentcolon= \alpha^*(\bSigma)$ and $ \widehat{\alpha}^* \vcentcolon= \alpha^*(\hbSigma)$ as finite-sample adjustments considering the true and estimated covariance matrix, respectively. 
	Our notation omits the dependence of $\alpha^*$ on $\nu$ and $\alpha$ since these quantities are known. 
	
	\begin{figure}
		\centering
		\includegraphics[width=1\textwidth]{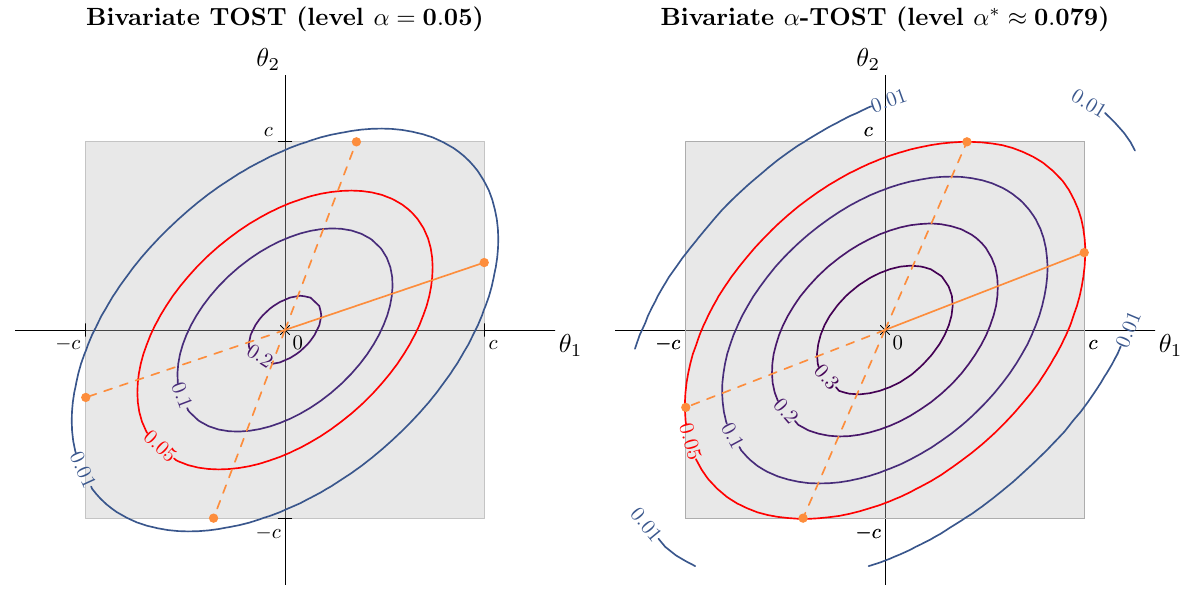}
		\caption{Target parameter space for the bivariate TOST (left panel) and bivariate $\alpha$-TOST (right panel) in a homoscedastic setting with $\sigma=0.1$, $\rho=0.8$, $c=\log(1.25)$ and $\alpha=0.05$. The outer square margins, taking the values $-c$ or $c$, correspond to the boundary between the spaces of the null and alternative hypotheses. The contour ellipses show the probability of rejecting the null hypothesis for a given value of $\btheta$. The ellipses in red correspond to a probability $\alpha$. Orange dots correspond to the coordinates of the set of suprema of the probability of declaring equivalence over the space of the null hypothesis. These coordinates are (slightly) different for both methods due to different test levels being used. The solid orange lines provide a one-dimensional projection allowing us to study the operating characteristics of both methods in Section \ref{sec:simulation}, and the dashed orange lines represent other projections that would lead to the same results. We can note that, in the scenario considered here, the bivariate TOST is quite conservative whilst the bivariate $\alpha$-TOST offers a solution with size $\alpha$.}
		\label{fig:size_contours}
	\end{figure}

	As in the univariate case, there are instances where $\alpha^*$ in \eqref{eq:alpha_TOST_pop} does not exist. Let us consider the conditions of existence of $\alpha^*$ under three scenarios of dependence between the $\htheta_j$ elements: independence, nearly perfect dependence, and in-between cases. Under independence, i.e.,~when $\bSigma$ is diagonal, it is required that
	\begin{equation}
		\label{eq:cond_existence}
		\sigma_{\max} < \frac{2c}{\Phi^{-1}\left(\alpha^{1/m}+1/2\right)},
	\end{equation}
	where $\sigma_{\max}\vcentcolon=\max_{j\in\{1,\dots,m\} }\sigma_j$ and $\alpha < (1/2)^m$, for a solution to exist (refer to Appendix~\ref{appendix:existence_alphaTOST} for further details related to the conditions for the existence of $\alpha^*$). This implies that, under independence and for the customary equivalence limits and significance level that are used by regulatory agencies (i.e.,~$c=\log(1.25) \approx 0.223$ and~$\alpha = 0.05$), the adjustment for the multivariate $\alpha$-TOST is guaranteed to exist for $m \leq  4 $ as long as $\sigma_{\max}< 0.231$. Note that independence between $\htheta_j$ elements is an unlikely scenario and that variances above 0.2 are fairly large in equivalence testing given the values of $c$ typically used. Under nearly perfect dependence, i.e.,~when the absolute correlations between the $\htheta_j$'s are close to 1, the multivariate problem becomes close to a univariate assessment of equivalence on the dimension with the largest variance, and the condition of existence of the adjusted level is of the same form as \eqref{eq:cond_existence} where $m$ is simply replaced by 1, reducing the conditions to a single criterion on $\sigma_{\max}$ for any number of dimensions $m$. Let $\rho_{ij}$ denote the correlation between $\htheta_i$ and $\htheta_j$. For $\alpha^*$ to exist when $|\rho_{ij}| \in (0,1)$, for all $i,j\in \{1,\dots,m\}$ and $i \neq j$,
	$\sigma_{\max}$ would need to satisfy a criterion situated between the conditions of existence related to the two extreme dependence scenarios discussed above. The dimension $m$ could therefore be much greater than 4 when the absolute correlations $|\rho_{ij}|$ are large. 
	Note that the conditions for the existence of $\widehat{\alpha}^*$ are equivalent when replacing $\sigma_{\max}$ by $\widehat{\sigma}_{\max}$ in condition $\eqref{eq:cond_existence}$.
	We believe this condition to be lenient and not to be respected in only rare and unusual cases. Note that, in the univariate context, similar requirements apply to other methods like the Reference-Scaled Bioequivalence,\cite{us2001fda} the Average Bioequivalence with Expanding Limits,\cite{committee2010ema} and their corrective procedures (see, e.g., Labes and Schütz\cite{labes2016inflation}). For a comprehensive comparison of these methods with the univariate $\alpha$-TOST, see Boulaguiem et al.\cite{boulaguiem23}
	
	By definition, when a solution to \eqref{eq:alpha_TOST_pop} exists, the resulting $\alpha^*$ guarantees a size of $\alpha$ for the multivariate equivalence test \eqref{eq:hypotheses_mult2} when used in \eqref{eqn:tstat-assess}. Figure~\ref{fig:size_contours} compares the test size of the conventional TOST and $\alpha$-TOST procedures, in a bivariate setting,  for the set of coordinates corresponding to the supremum of the probability of declaring equivalence over the space of the null hypothesis. In the scenario considered here, the bivariate TOST is quite conservative whilst the bivariate $\alpha$-TOST has a solution and is size $\alpha$ when $\alpha^* \approx 0.079$ is used as test level. Note that \eqref{eq:alpha_TOST_pop} implies the use of the same corrected significance level to construct each $100(1-2\alpha^{*})$\% marginal confidence interval. 
	
	The asymptotic properties of $\widehat{\alpha}^*$ are investigated in Appendix~\ref{appendix:asymptotic_alphaTOST}. We establish that, for any $m\geq1$, $\widehat{\alpha}^*$ converges to ${\alpha}^*$ at a rate of $o_p\left(\nu^{-1}\right)$. This result suggests that the asymptotic uncertainty associated with $\widehat{\alpha}^*$ becomes negligibly small in comparison to the uncertainties associated with $\hbtheta$ and $\hsigma_j$, for all $j\in\{1,\dots,m\}$, which exhibit slower convergence rates of $\mathcal{O}_p(\nu^{-1/2})$ and $\mathcal{O}_p(\nu^{-1})$, respectively.
	In practice, although the empirical size is not guaranteed to be $\alpha$, it is therefore reasonable to expect very similar finite-sample behaviour for the multivariate $\alpha$-TOST procedures based on $\alpha^*$ or $\widehat{\alpha}^*$.
	Importantly, our simulation results presented in Section~\ref{sec:simulation} show that the empirical size of the multivariate $\alpha$-TOST based on $\widehat{\alpha}^*$ remains below the target nominal level $\alpha$, indicating that it effectively controls the type I error.
	
	Finally, since the multivariate $\alpha$-TOST provides an adjusted level $\alpha^*(\bOmega) \geq \alpha$ (with equality holding only in the limiting case $\sigma_{\max} \to 0$), its rejection region cannot be smaller than that of the multivariate TOST, which makes the former uniformly more powerful. This is reflected in the case study presented in Section~\ref{sec:application}, where the application of the multivariate $\alpha$-TOST results in the shrinkage of the confidence intervals enabling it to declare equivalence, a determination that the conventional multivariate TOST fails to achieve.
	
	\subsection{Computational Details}
	\label{subsec:comput}
	
\begin{figure}[t!]
		\centering
		\includegraphics[width=1\textwidth]{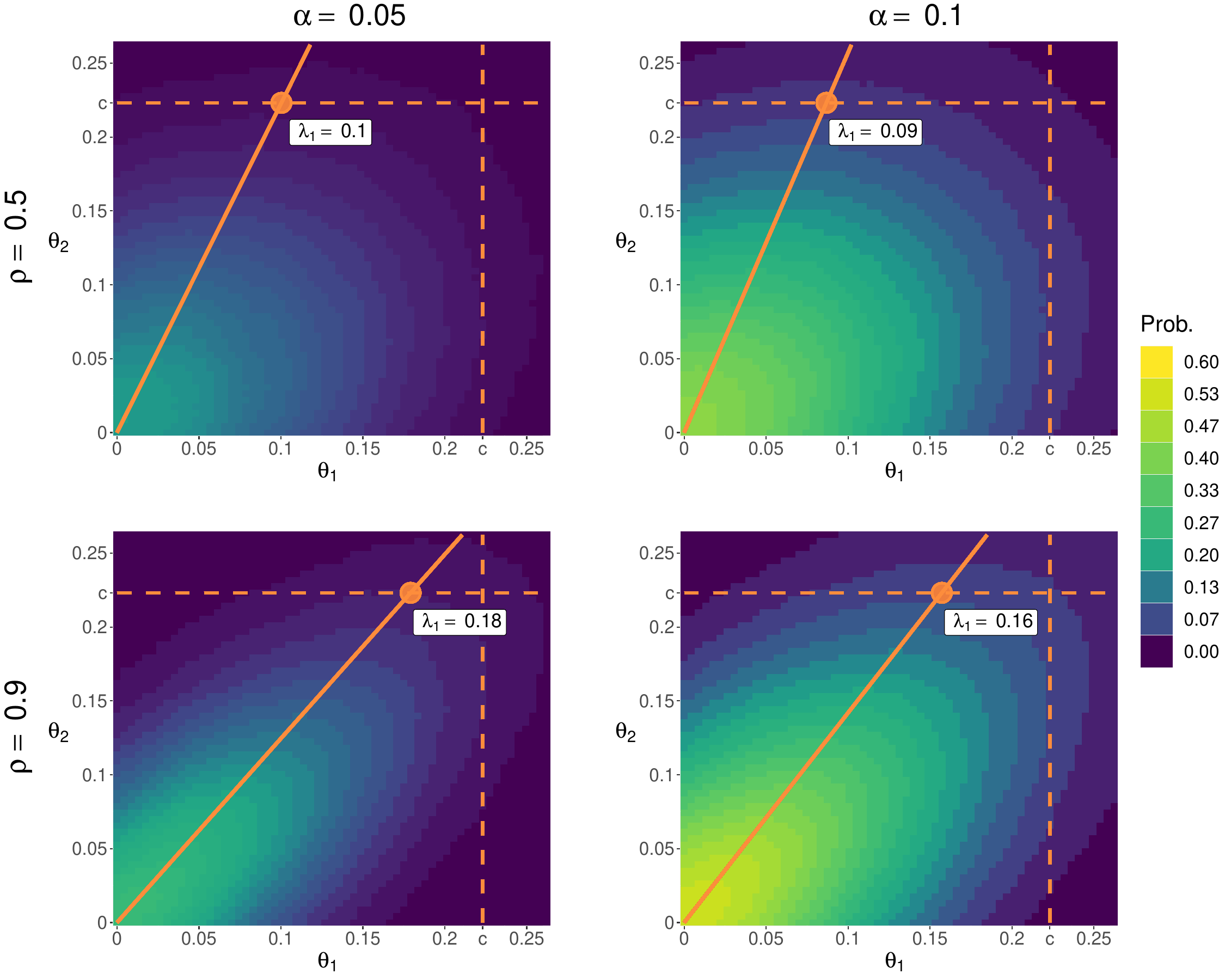}
		\caption{Colour-coded probabilities of the bivariate TOST rejecting $\text{H}_0$ for each coordinate of the upper right quadrant of the target parameter space in four settings (panels) defined as a combination of two dependence levels between the $\htheta_j$ elements (rows) and two test levels (columns) when $\sigma_1=\sigma_2=0.1$, $\nu=20$, $c=\log(1.25)$. The orange dot corresponds to $\blambda $ in \eqref{eq:lambda}, the orange dashed lines represent the boundary of the rejection region of the bivariate TOST, and the orange solid line represents the linear combination $\kappa \blambda$ on which the operating characteristics of the procedure are defined in Section \ref{sec:simulation}, with $\kappa = [0, 1.2]$.}
		\label{fig:contour-lambda}
	\end{figure}

	Compared to the univariate case, solving~\eqref{eq:alpha_TOST_pop} is far more challenging in the multivariate setting since $\blambda(\gamma)$ typically depends on the significance level $\gamma$, over which the optimisation is performed. As an illustration, Figure \ref{fig:contour-lambda} shows the colour-coded probabilities of the bivariate TOST rejecting $\text{H}_0$ for each coordinate of the upper right quadrant of the target parameter space in four settings (panels) defined as combination of two dependence levels between the $\htheta_j$ elements (rows) and two test levels (columns) when $\sigma_1=\sigma_2=0.1$, $\nu=20$, $c=\log(1.25)$. In each panel, the orange dot shows the $\blambda$-coordinates corresponding to the supremum of the probability of rejecting $\text{H}_0$ over the space of the null hypothesis in \eqref{eq:size}.
	This supremum is always located on the boundary of the null space, represented on the graph by dashed orange lines.
	This graph illustrates that $\blambda$ depends on both the covariance structure and the test level. Indeed, we can note that $\blambda$ is not located at the same $(\lambda_1, c)$ coordinates for the different correlation and test levels, showing the difficulty of defining $\alpha^*$ in the multivariate setting, as this requires a recursive optimisation in which $\alpha^*$ is optimised for a given $\blambda(\alpha)$, from which $\blambda(\alpha^*)$ needs to be re-computed until convergence.

	Our recursive and iterative procedure to compute $\alpha^*$ in \eqref{eq:alpha_TOST_pop} is described in Algorithm~\ref{alg:multiv_alphaTOST}, and a formal description of our algorithm is provided in Appendix~\ref{appendix:uniqueness_alphaTOST}. Our empirical experience suggests that this procedure is computationally lean and stable. Numerically, our algorithm takes advantage of the fact that, for a fixed $\blambda$, which corresponds to the outer loop indexed by $r\in\mathbb{N}$, the adjusted significance level $\alpha^{*(r)}$ in the inner loop can be computed following a simple iterative approach that converges exponentially fast. Indeed, at iteration $k\geq1$, we have
	\begin{equation}
		\alpha^{(r)}_k = \alpha^{(r)}_{k-1} + \alpha - p\left[ \alpha^{(r)}_{k-1}, \blambda\left\{\alpha_0^{(r)}\right\}, \bOmega, \nu , \bm{c} \right],
	\end{equation}
	where $\alpha_0^{(0)}=\alpha$, and $\alpha_0^{(r)}=\alpha^{*(r-1)}$ for $r\geq 1$ (which is the solution at the previous iteration of the outer loop). We show in Appendix~\ref{appendix:uniqueness_alphaTOST} that this iterative approach converges exponentially fast to a unique solution $\alpha^{*(r)}$ under mild conditions which are usually satisfied when $\sigma_{\max}$ is not large compared to $c$. In particular, we show that there exists a constant $b>0$ such that
	\begin{equation*}
		\Big| \alpha^{(r)}_k  - \alpha^{*(r)}  \Big| < \frac{1}{2} \exp(-bk).
	\end{equation*}
	The logic remains the same for finite-sample adjustments where the population quantities can simply be replaced by their empirical counterparts. 
	
	Compared to the conventional TOST, constructing the multivariate $\alpha$-TOST adds some computational burden due to the estimation of $\alpha^*$ via numerical methods, but the current implementation remains fairly computationally efficient. For instance, in realistic applications such as the one considered in Section~\ref{sec:application}, where $\bSigma$ is unstructured and $m=4$, getting the multivariate $\alpha$-TOST results requires
	between 1 to 10 seconds on a standard computer depending on the precision of the numerical method (e.g.,~Monte Carlo integration) used to approximate the probability of rejecting H$_0$ in \eqref{eq:TOST_power}.

	\vspace{0.5cm}
	\begin{algorithm}[H] \label{alg:multiv_alphaTOST}
		\SetAlgoLined
		\Input{nominal significance level $\alpha$, covariance matrix $\bOmega$, degrees of freedom $\nu$, equivalence margins $\bm{c}$, maximum number of iterations $r_{\max}$, algorithmic tolerance $\epsilon_{min}$; }
		\Output{adjusted significance level $\alpha^*$; } 
		Initialise $\alpha^{*(0)} = \alpha$, compute $\blambda(\alpha^{*(0)})$, and set $r = 0$\;
		\While{
			$ \left\lvert p \left[ \alpha^{*(r)}, \blambda \{ \alpha^{*(r)} \}, \bOmega, \nu , \bm{c} \right ] -\alpha \right\rvert > \epsilon_{\min}$ and $r \leq r_{\max}$ }{
			Initialise $\alpha^{(r)}_{0} = \alpha^{*(r)}$, compute $\alpha^{(r)}_1 $ and set $k = 1$\;
			\While{ $ \lvert \alpha^{(r)}_k - \alpha^{(r)}_{k-1} \rvert > \epsilon_{\min}$ }
			{$k = k+1$\;
				Compute $ \alpha^{(r)}_k $;
			}
			$r = r+1$\;
			Store the adjusted level $\alpha^{*(r)} = \alpha^{(r-1)}_{k}$\;
			Compute $\blambda(\alpha^{*(r)}) $;
		}             
		\caption{Algorithm for computing the adjusted level $\alpha^{*}$ for the multivariate $\alpha$-TOST}
	\end{algorithm}
	\vspace{0.5cm}

	\section{Simulation Study}
	\label{sec:simulation}
	
	In this section, we compare the operating characteristics of the conventional and $\alpha$- multivariate TOST procedures by means of an extensive simulation study encompassing the four settings described in Table~\ref{fig:sim_table}. We considered settings with small to medium sample sizes ($\nu \in \{20, 40\}$), numbers of outcomes typical in bioequivalence studies ($m \in \{2,4\}$), a wide range of dependence levels between the $\htheta$ elements ($\rho \in \{0, 0.5, 0.9\}$), small to large variance levels for the $\htheta_j$ ($\sigma_j \in \{0.05, 0.1, 0.15\}$), and different variance (both homo- and heteroscedastic cases) and covariance (compound symmetry and AR1 dependence) structures, with $c=\log(1.25)$, $\alpha=0.05$ and a large number of Monte Carlo samples ($B = 5 \times 10^4$ per scenario). 
	
	\begin{table}[]
		\centering
		\includegraphics[scale=0.8]{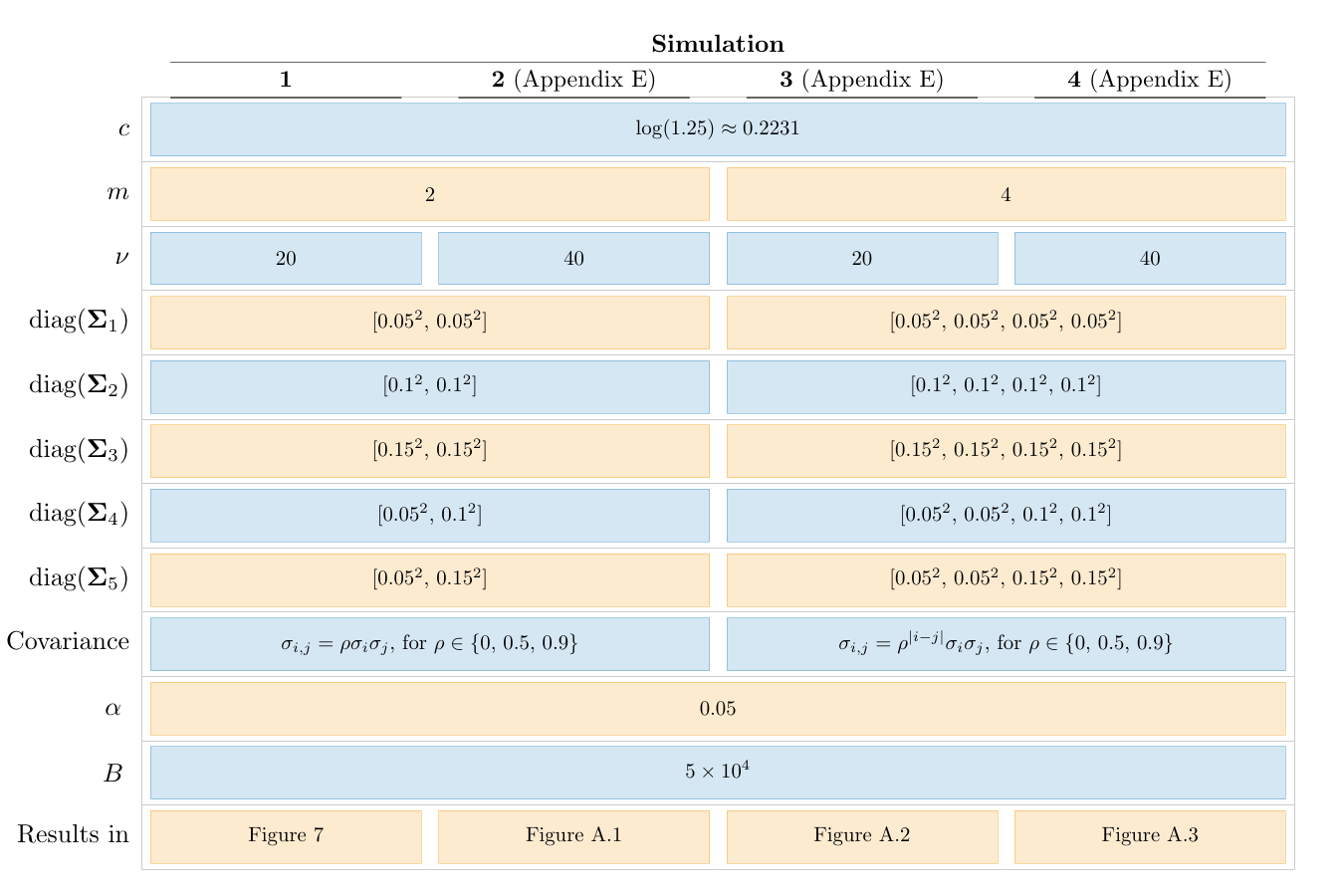}
		\caption{Parameter values used in each setting of the simulation, where $c$ denotes the tolerance limit, $m$ the number of outcomes, $\nu$ the number of degrees of freedom, $\bm{\Sigma}$ the covariance matrix of $\hbtheta$ with $\sigma_{i,j}$ as off-diagonal element, $\alpha$ the target significance level and $B$ the number of Monte Carlo samples per simulation.}
		\label{fig:sim_table}
	\end{table}
	
\begin{figure}[t!]
		\centering
		\includegraphics[width=0.5\linewidth]{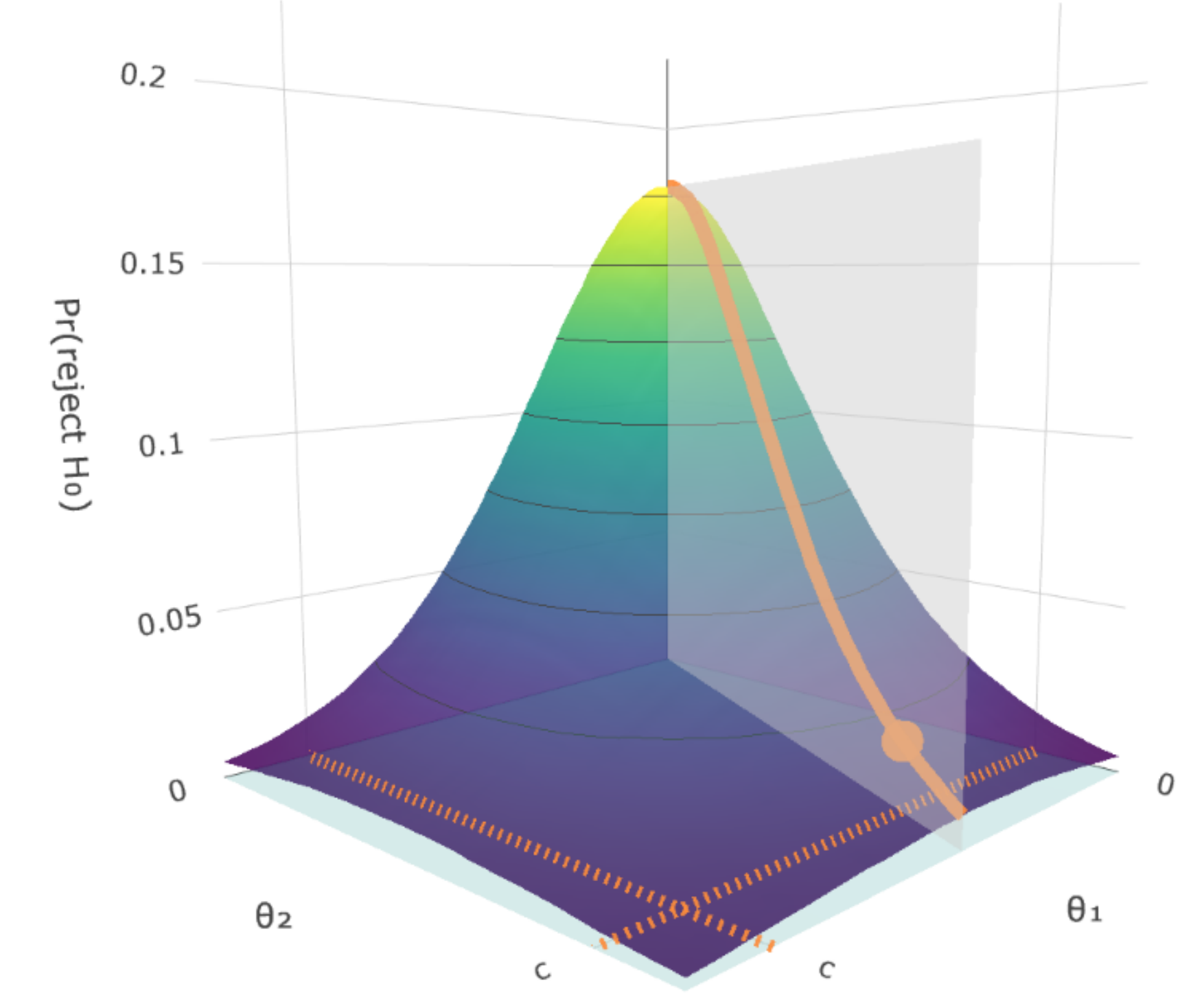}
		\caption{3D-heatmap of the probabilities of rejecting the null hypothesis (z-axis) in a bivariate equivalence case (x- and y- axes). The combinations of the target parameter values $\kappa \blambda$ used to compare different methods correspond to the projection of the light grey plane on the horizontal plane. 
			The associated probabilities (of rejecting the null hypothesis) are displayed with the orange solid line and the orange dot shows the test size at coordinates $\blambda = (\lambda_1, c)$, which is obtained for $\kappa=1$.}
		\label{fig:3d_plot}
	\end{figure}

	Our simulations consider the canonical setting
	\begin{equation} \label{eqn:canon-kappa}
		\widehat{\bm{\theta}} \sim \mathcal{N}_m \left\{ \kappa \blambda ,\bm{\Sigma} \right \} \quad  \text{and} \quad 
		\nu \hbSigma \sim \mathcal{W}_m \left(\nu, \bm{\Sigma} \right),   
	\end{equation} 
	where $\kappa \in [0, 1.2]$, and $\blambda$ represents the coordinates of the parameter space under the null hypothesis that determine the test size, as defined in \eqref{eq:lambda}. $\kappa\blambda$ consists in a linear combination that includes both $0$ and $\blambda$, thus allowing us to define the probability of declaring equivalence both under the alternative hypothesis for $\kappa\in [0,1)$ (representing the power), and under the null hypothesis for $\kappa\in [1,1.2]$, where the test size is evaluated at $\kappa=1$.
	As each method has its own test level dependent $\blambda$ coordinates, as shown in \eqref{eq:lambda}, the trajectories spanned by $\kappa\blambda$ allow for a valid comparison between methods. We therefore consider for the simulation study 30 equally-spaced values of $\kappa\in[0,1.2]$. In the 3D-heatmap of the probabilities of rejecting the null hypothesis (i.e., declaring equivalence) in a bivariate equivalence case shown in Figure \ref{fig:3d_plot}, the $\kappa \blambda$ set of values corresponds to the projection of the light grey plane on the horizontal plane. 
	Therefore, for $\kappa=0$ we obtain the coordinates $(0,0)$ and for $\kappa=1$ we obtain $\blambda = (\lambda_1, c)$, corresponding to the coordinates of the orange dot in the horizontal plane. 
	The associated probabilities (of rejecting the null hypothesis) are displayed with the orange solid line and the orange dot shows the test size at coordinates $\blambda = (\lambda_1, c)$.

\begin{figure*}[ht]
		\centering
		\includegraphics[width=0.75\textwidth]{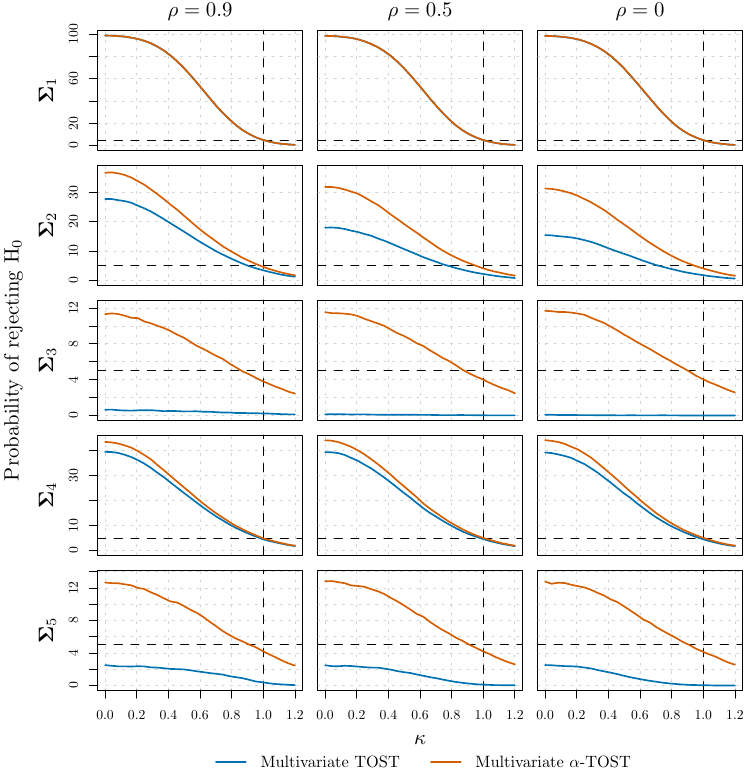}
		\caption{Probability of rejecting the null hypothesis (y-axis) as a function of $\kappa \in [0, 1.2]$ (x-axis) for different sets of variances (rows) and levels of dependence (columns) for the conventional multivariate TOST (blue line) and the multivariate $\alpha$-TOST (orange line) in the settings considered in Simulation 1 (refer to the first column of Table~\ref{fig:sim_table} for details).}
		\label{fig:sim1}
	\end{figure*}

	Figure~\ref{fig:sim1} shows the results of Simulation 1. Each panel displays the probability of rejecting the null hypothesis (y-axis) as a function of $\kappa \in [0, 1.2]$ (x-axis) for different sets of variances (rows) and levels of dependence (columns) for the conventional multivariate TOST (blue line) and the multivariate $\alpha$-TOST (orange line).
	\par We can note that, with equal and small variances for the $\htheta_j$ elements ($\bm{\Sigma}_1$), the probability of rejecting $\text{H}_0$ of both methods overlap. In such settings indeed, the finite-sample adjustments of multivariate $\alpha$-TOST are negligible due to the multivariate TOST being almost size-$\alpha$, irrespective of the level of dependence between elements of $\hbtheta$.
	With equal and moderate to large variances for the $\htheta_j$ elements however ($\bm{\Sigma}_2$ and $\bm{\Sigma}_3$), the empirical size of the multivariate TOST ($\kappa=1$) strongly decreases when the variances of the $\htheta_j$ increase and when the dependence level decreases. In such cases, its empirical size is considerably smaller than the nominal level of $\alpha=0.05$ and often reaches 0 ($\bm{\Sigma}_3$).
	The multivariate $\alpha$-TOST on the other hand shows an empirical size that is close to the nominal significance level, albeit slightly conservative for larger variances, leading to a higher probability of rejecting $\text{H}_0$ for all values of $\kappa$, i.e., to a higher power ($\kappa < 1$). Specifically, for $\sigma_1 = \sigma_2=0.1$ and $\rho=0.9$, the multivariate $\alpha$-TOST reports an empirical power at $\kappa=0$ which is 10\% larger than the one of the multivariate TOST, and this gap widens for smaller dependence levels. We can note that the probability of rejecting $\text{H}_0$ across $\kappa$'s decreases with $\rho$ for both methods, albeit much more markedly for the multivariate TOST. For any dependence level, the homoscedastic scenarios leading to the operating characteristics 
	that are most similar to the ones of heteroscedastic scenarios ($\bm{\Sigma}_4$ and $\bm{\Sigma}_5$) are the scenarios corresponding to the largest variance of the elements $\hbtheta$ (i.e., $\bm{\Sigma}_2$ for $\bm{\Sigma}_4$, and $\bm{\Sigma}_3$ for $\bm{\Sigma}_5$). 
	Overall, the multivariate $\alpha$-TOST is therefore expected to offer substantial gains in settings where at least one of the variances is not too small, as its operating characteristics are mainly driven by the largest variance.
	
	The conclusions of our 3 other simulation settings, considering other sample sizes ($\nu = 40$ in Simulations 2 and 4) and dependence patterns (AR1 dependence type in Simulations 3 and 4), are very similar to the ones discussed here, thus showing 
	that
	the multivariate finite-sample adjustment of our method effectively improves the properties of the conventional multivariate TOST in virtually all cases not considering small variances for the $\htheta_j$. 
	
	We remark that the size of multivariate $\alpha$-TOST is guaranteed to be $\alpha$ only at the population level when 
	an $\alpha^*$ adjustment is used. 
	However, these simulation results show that its feasible counterpart, based on $\widehat{\alpha}^*$, leads to an empirical size that remains below the nominal level $\alpha$.
	Therefore, the procedure effectively controls the type I error.
	This conservative behaviour has also been documented for the univariate $\alpha$-TOST procedure\cite{boulaguiem23}, where the phenomenon was investigated in greater depth, and we conjecture that it extends to the multivariate framework.

	\section{Case Study: Ticlopidine Hydrochloride}
	\label{sec:application}
	
	In this section, we re-visit the case study on ticlopidine hydrochloride of Marzo et al.\cite{marzo2002bioequivalence} and compare the results of the conventional and $\alpha$- multivariate TOST procedures when simultaneously assessing bioequivalence for four pharmacokinetic parameters.
	
	Ticlopidine hydrochloride (CAS 55142-85-3) is an inhibitor of platelet aggregation known for its effective antithrombotic activity. It is therapeutically used to manage thromboembolic disorders, especially in high-risk patients for stroke and myocardial infarction. Marzo et al.\cite{marzo2002bioequivalence} assessed the bioequivalence of a new formulation of ticlopidine hydrochloride (referred to as $T$) with the formulation that was marketed at that time, called Tiklid (referred to as $R$). The study involved $n=24$ healthy male volunteers who received both formulations in the form of a tablet containing \SI{250}{\mg} of active ingredient in a $2\times2\times2$ crossover design with a washout period of three weeks between the two administrations. The purpose of this assessment was to register the new formulation as a generic drug, following the Abbreviated New Drug Application procedure. 
	
	The original pharmacokinetic analysis ($m=5$) included the evaluation of the C$_{\max}$ (maximum plasma concentration), t$_{\max}$ (time to reach C$_{\max}$), AUC$_{0-t}$ (area under the concentration-time curve from time zero to the last measurable concentration), AUC$_{0-\infty}$ (area under the concentration-time curve from time zero to infinity), and t$_{1/2}$ (elimination half-life) using non-compartmental methods. 
	We omitted
	t$_{\max}$ from our analysis due to its discrete nature, reducing the number of variables to $m=4$. 
	We chose to also include  t$_{1/2}$ in our analysis, 
	although Food and Drug Administration’s guidance only requires C$_{\max}$, AUC$_{0-t}$ and AUC$_{0-\infty}$ to be assessed for  bioequivalence.\cite{pallmann2017simultaneous} We pre-processed the data to remove evident outliers in t$_{1/2}$, bringing the available sample size to $n=20$. Additional details on data pre-processing can be found in Appendix~\ref{app:application}. 
	\begin{figure}[ht]
		\centering
		\includegraphics[width=0.9\textwidth]{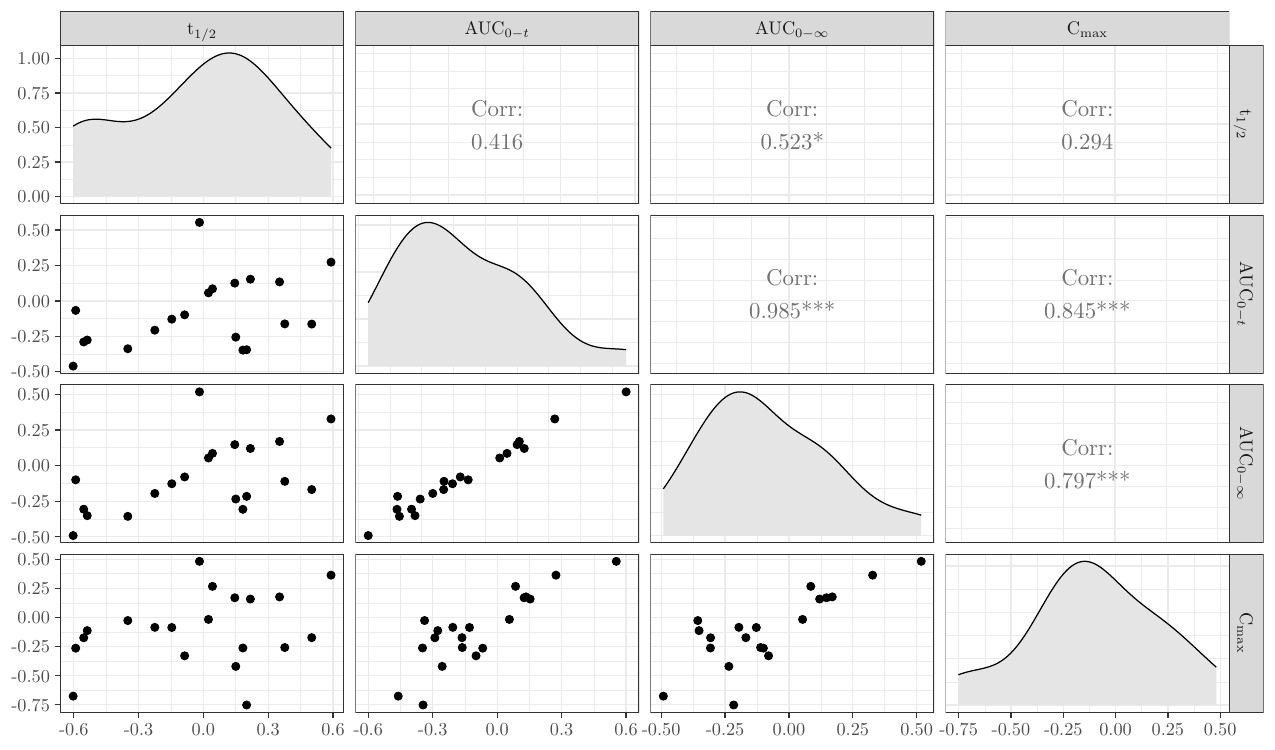}
		\caption{Scatterplots, kernel density estimates and correlations of the (logarithmically transformed) differences between the two formulations of ticlopidine hydrochloride for each pair of the pharmacokinetic measurements. The symbols $^{***}$, $^{**}$ and $^{*}$ indicate $p$-values smaller than 0.001, 0.01 and 0.05, respectively, for testing H$_0: \rho_{i,j} = 0$ vs. H$_1: \rho_{i,j} \neq 0$, where $\rho_{i,j}$ corresponds to the correlation between the variables in row $i$ and column $j$.  The figure was obtained using the default parameters of the \texttt{ggpairs} function from the \texttt{GGally} package in \texttt{R}.\cite{ggpairs}}
		\label{fig:pairscatter_subset}
	\end{figure}
	
	Figure~\ref{fig:pairscatter_subset} shows scatterplots of the differences between the two formulations ($T-R$) for each pair of pharmacokinetic outcomes after applying the logarithmic transformation, as well as their pairwise Pearson correlations and the associated $p$-values.
	The target parameters related to C$_{\max}$, AUC$_{0-\infty}$ and AUC$_{0-t}$ show strong positive correlation levels whilst the target parameter related to t$_{1/2}$ is more loosely related with the other ones. Since $m=4$ and $\widehat{\sigma}_{\max}=0.082<0.231$ 
	(which corresponds to the standard error of t$_{1/2}$), the sufficient criteria on the existence of a solution for the multivariate $\alpha$-TOST under independence are satisfied, which guarantee the existence of a solution here as the present data show strong dependence levels.
	
\begin{figure}[ht]
		\centering
		\includegraphics[width=0.645\linewidth]{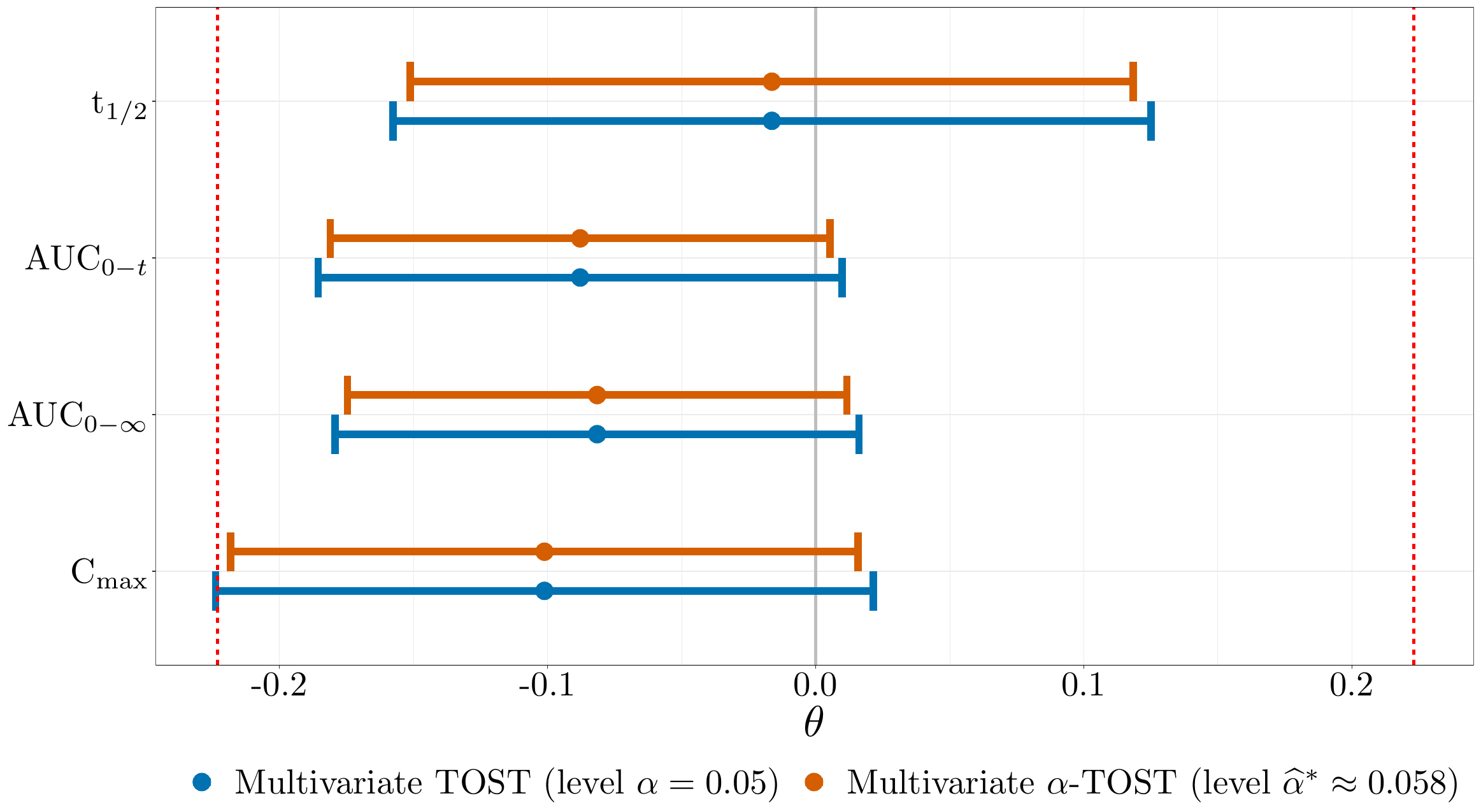}
		\caption{Intervals showing the target parameter 100$(1-2\alpha)$\% confidence intervals of both methods (colours) for each outcome (rows)  on the logarithmic scale, with $\alpha=0.05$. 
			Marginal equivalence bounds $(-c,c)$, with $c\approx 0.223$, are represented by dashed vertical lines in red.
			Confidence interval limits are reported in Table~\ref{tab:app_ci}.}
		\label{fig:app_ci}
	\end{figure}

	Table~\ref{tab:app_ci} and Figure~\ref{fig:app_ci} respectively report and display the $100(1-2\alpha)$\% marginal confidence intervals defined on the logarithmic scale for the target parameter related to each pharmacokinetic outcome for both methods. Based on the interval-inclusion principle \cite{wellek2010testing} and using the symmetric standard equivalence margins represented by $c = \log(1.25) \approx 0.223$ and nominal significance level $\alpha=0.05$, the multivariate TOST cannot declare
	bioequivalence since the confidence interval of C$_{\max}$ exceeds the equivalence margins. On the other hand, our multivariate $\alpha$-TOST procedure is able to declare the equivalence of the two formulations due to a larger test level ($\widehat{\alpha}^* \approx 0.058$) being used to reach a test size of 5\%, leading to narrower confidence intervals.
	\begin{table}[ht]
		\centering
		\begin{tabular}{lccccc}
			\toprule
			Method & t$_{1/2}$ & AUC$_{0-t}$ & AUC$_{0-\infty}$ & C$_{\max}$ & Bioequivalence \\
			&   &   &   &   & Declaration \\
			\midrule
			Multivariate TOST & $(-0.158, 0.125)$ & $(-0.186, 0.010)$ & $(-0.179, 0.016)$ & $(-0.224, 0.022)$ & No \\  
			Multivariate $\alpha$-TOST  & $(-0.151, 0.118)$ & $(-0.181, 0.005)$ & $(-0.175, 0.012)$ & $(-0.218, 0.016)$ & Yes \\ 
			\bottomrule
		\end{tabular}
		\caption{Target parameter 100$(1-2\alpha)$\% confidence intervals of both methods (rows) for each outcome (columns) on the logarithmic scale with $\alpha=0.05$. 
			Marginal equivalence bounds are $(-c,c)$ with $c\approx 0.223$.}
		\label{tab:app_ci}
	\end{table}
	\section{Final Remarks}
	\label{sec:conclusion}
	
	We investigated finite-sample adjustments for multivariate average equivalence testing using the very general canonical framework in \eqref{eq:canon} covering most designs used in this setting.
	We demonstrated the theoretical and finite-sample properties of the proposed approach, leading to a procedure that controls the test size and is uniformly more powerful than the conventional multivariate TOST. Our procedure is therefore particularly useful in the presence of relatively large standard errors, like in the case of highly variable drugs for example, where the conventional multivariate TOST is known to be highly conservative.
	
	As it is the case with the multivariate TOST, the multivariate $\alpha$-TOST might be influenced by extreme observations, both when estimating the mean responses as well as the covariance structure. The effect of such observations on the resulting estimators, as well as the testing procedure itself, is still unclear, but we can conjecture that the size of the tests might be biased, leading to a potential loss of power. Investigating the effects of extreme observations and proposing a robust alternative procedure is certainly worth doing, but this is left for further research.
	
	The proposed approach could also be extended to a variety of other settings. The inclusion of covariates could allow reducing the residual variance. Extensions to non-linear cases, such as to
	binary outcomes, for example, would follow the same logic, but would require a specific treatment due to the nature of the responses and to the use of link functions. These exciting extensions are deferred to further research.

	\cleardoublepage
	\newpage
	\bibliographystyle{unsrtnat-initials} 
	\bibliography{refs}
	
	\clearpage
	\newpage
	
	\appendix
	
	\section*{\huge{Appendix}}
	\label{sec:app}
	
	\renewcommand{\thetable}{A.\arabic{table}}
	\setcounter{table}{0}
	\renewcommand{\theequation}{A.\arabic{equation}}
	\setcounter{equation}{0}
	\renewcommand{\thefigure}{A.\arabic{figure}}
	\setcounter{figure}{0}

	\section{Asymptotic Size of the Multivariate TOST}
	\label{appendix:size_TOST}
	
	Here we demonstrate that the multivariate TOST is size-$\alpha$ asymptotically, namely that
	$$
	\lim_{\substack{\nu \to \infty}}  
	p ( \alpha, \blambda, \bSigma, \nu , \bm{c} )
	= \alpha.
	$$
	
	Notice that, given $\alpha$, $\blambda$ (as defined in Equation \ref{eq:lambda}), $\bSigma$, $\nu$ and $\bm{c}$, the size can be written as follows
	\begin{equation}
		\begin{aligned}
			p(\alpha, \blambda, \bSigma, \nu , \bm{c} ) &= 
			\Pr \left(  
			\bigcap_{j=1}^m \left\{ \lvert \htheta_j \rvert < c - t_{\alpha,\nu}  \hsigma_{j} \right\}
			\right) \\
			&= \Pr \left(  
			\bigcap_{j=1}^m \left\{ -\frac{\lambda_j+c}{\hsigma_j} + t_{\alpha,\nu}\leq T_{\nu,j}\leq -\frac{\lambda_j-c}{\hsigma_j} - t_{\alpha,\nu} \right\} 
			\right),
		\end{aligned}
		\label{eq:joint:tost}
	\end{equation}
	where $T_{\nu,j}$ follows a Student's $t$-distribution with $\nu$ degrees of freedom. If $\max\limits_{i \neq j} \, \lvert\rho_{i,j}\lvert \, <1$ where $\rho_{ij}$ corresponds to the correlation between the $i$-th and $j$-th components, the vector $\blambda$ contains $m-1$ elements between $(-c, c)$ and a single element at $c$. Without loss of generality, let us consider
	$\lambda_1 = c$. Therefore, we can rewrite the joint distribution in \eqref{eq:joint:tost} as follows: 
	\begin{equation*}
		\begin{aligned}
			&\Pr\left(  
			\bigcap_{j= 1}^m \left\{ -\frac{\lambda_j+c}{\hsigma_j} + t_{\alpha,\nu}\leq T_{\nu,j}\leq -\frac{\lambda_j-c}{\hsigma_j} - t_{\alpha,\nu} \right\} 
			\right)\\ 
			&=\Pr ( E )  \Pr\left( \bigcap_{\substack{j =2}}^m \left\{ -\frac{\lambda_j+c}{\hsigma_j} + t_{\alpha,\nu}\leq T_{\nu,j}\leq -\frac{\lambda_j-c}{\hsigma_j} - t_{\alpha,\nu} \right\}
			\lvert 
			E 
			\right)
		\end{aligned}
	\end{equation*}
	where $E \vcentcolon= \left\{-\frac{2c}{\hsigma_1}+t_{\alpha,\nu}\leq T_{1,\nu}\leq -t_{\alpha,\nu}\right\}$. Consequently, asymptotically we get 
	\begin{equation*}
		\begin{aligned}
			\lim_{\nu \to \infty} 
			&\Pr \left( E \right) = \Pr \left(Z \leq -z_{\alpha}\right)=\alpha \\
			\lim_{\nu \to \infty} & \Pr \left( \bigcap_{\substack{j =2}}^m \left\{ -\frac{\lambda_j+c}{\hsigma_j} + t_{\alpha,\nu}\leq T_{\nu,j}\leq -\frac{\lambda_j-c}{\hsigma_j} - t_{\alpha,\nu} \right\}
			\Big\lvert E \right) = 1,
		\end{aligned}
	\end{equation*}
	where $Z$ follows a standard normal distribution,  and finally
	\begin{equation}
		\begin{aligned}
			\lim_{\substack{\nu \to \infty}} p ( \alpha, \blambda, \bSigma, \nu , \bm{c} ) = \alpha .
		\end{aligned}
		\label{eq:asympsize:tost}
	\end{equation}
	
	\section{Derivations of a simplified TOST procedure}
	\label{appendix:simple_TOST}
	
	We consider a simplified $m$-dimensional setting (with $m > 1$) for the canonical form in \eqref{eq:canon}. Namely:
	\begin{equation} 
		\widehat{\bm{\theta}} \sim \mathcal{N}\left(\bm{\theta}, \bm{\Sigma}\right), \nonumber
	\end{equation} 
	for a known $\bSigma  = \sigma^2 \mathbf{I}_m$.
	The multivariate TOST procedure corresponds to $m$ univariate TOST procedures. Therefore, we define:
	\begin{equation}
		Z_L^{(j)} \vcentcolon=\frac{\widehat{{\theta}}_j + c}{{\sigma}} \quad \text{and} \quad  Z_U^{(j)} \vcentcolon=\frac{\widehat{{\theta}}_j - c}{{\sigma}},
	\end{equation}
	for $j = 1, \ldots, m$ and where $\widehat{\theta}_j$ denotes $j$-th entry of $\widehat{\bm{\theta}}$. 
	We declare equivalence (i.e.,~reject H$_0$) if
	\begin{equation}
		\min_{j = 1, \ldots, m} \, Z_L^{(j)} \geq z_{\alpha} \;\;\; \text{{and}} \quad \max_{j = 1, \ldots, m} \, Z_U^{(j)} \leq -z_{\alpha} ,
	\end{equation}
	where $z_{\alpha}$ denotes the upper $\alpha$ quantile of a standard normal distribution.
	The probability of accepting the alternative hypothesis is given by
	\begin{equation}
		\begin{aligned}
			p (\bm{\theta}) \vcentcolon=& \Pr\left( \min_{j = 1, \ldots, m} \, Z_L^{(j)} \geq z_{\alpha} \; \text{and}\;\; \max_{j = 1, \ldots, m} \, Z_U^{(j)} \leq -z_{\alpha} \mid \btheta, \sigma, \alpha, c, m \right)\\
			=& \prod_{j = 1}^m \Pr\left(\frac{\widehat{\theta}_j + c}{{\sigma}} \geq z_{\alpha}   \; \text{and}\;\; \frac{\widehat{\theta}_j - c}{{\sigma}} \leq -z_{\alpha} \right)\\
			=& \prod_{j = 1}^m  \left\{1 - \Phi\left( z_{\alpha}  - \frac{c - {\theta}_j}{{\sigma}}\right) - \Phi\left( z_{\alpha}  - \frac{c + {\theta}_j}{{\sigma}}\right)\right\} .
		\end{aligned}
	\end{equation}
	Furthermore, for $i \in \{1, \ldots, m\}$, we have
	\begin{equation}
		\frac{\partial}{\partial \theta_i}\, p(\bm{\theta}) = \frac{1}{\sigma} \prod_{\substack{j=1 \\ j\neq i}}^p  \left\{1 - \Phi\left( z_{\alpha}  - \frac{c - {\theta}_j}{{\sigma}}\right) - \Phi\left( z_{\alpha}  - \frac{c + {\theta}_j}{{\sigma}}\right)\right\}
		\left\{\varphi\left( z_{\alpha}  - \frac{c + {\theta}_i}{{\sigma}}\right) - \varphi\left( z_{\alpha}  - \frac{c - {\theta}_i}{{\sigma}}\right) \right\},
	\end{equation}
	where $\varphi$ corresponds to the probability density function of a standard normal distribution.
	Similar to the univariate case, we have
	\begin{equation}
		\frac{\partial}{\partial \theta_i}\, p(\bm{\theta}) < 0,
	\end{equation}
	for all $i \in \{1, \ldots, m\}$. Moreover, the function $p(\bm{\theta})$ is both even, i.e., $p(\bm{\theta}) = p(-\bm{\theta})$ and symmetric, i.e., its value is the same no matter the order of its arguments. Thus, the size of the TOST procedure is given by
	\begin{equation}
		\begin{aligned}
			\sup_{\btheta \in \real^m \setminus \bm{\Theta}_1} \; p(\bm{\theta}) =
			\sup_{\btheta \in \real_+^m \setminus [0, c)^m} \; p(\bm{\theta}) =
			\sup_{\theta_1 \geq c, \theta_2 \geq 0, \ldots, \theta_m \geq 0} \; p(\bm{\theta}) = p(\bm{\lambda}),
		\end{aligned}
	\end{equation}
	where $\bm{\lambda} \vcentcolon= [c, 0, \ldots, 0]^T$. 
	Therefore, we obtain
	\begin{equation}
		\begin{aligned}
			p \{ \alpha, \blambda, \sigma^2\mathbf{I}_m , \bm{c} \} =
			\left\{1 - \Phi\left( z_{\alpha}\right) - \Phi\left( z_{\alpha}  - \frac{2 c }{{\sigma}}\right)\right\}
			\left\{1 - 2\Phi\left( z_{\alpha}  - \frac{c}{{\sigma}}\right)\right\}^{m-1}  ,
		\end{aligned}
	\end{equation}
	which completes the derivation of \eqref{eq:size_multiv_tost}.

	\section{Conditions on the Existence of the Multivariate $\alpha$-TOST}
	\label{appendix:existence_alphaTOST}
	
	In what follows, we discuss the conditions under which a solution of $\eqref{eq:alpha_TOST_pop}$ exists under independence. We fix $\alpha$, $\blambda$, $\bSigma$, $\nu$ and $\bm{c}$, and simplify the notation such that $p(\gamma)\vcentcolon=p \{ \alpha, \blambda(\gamma), \bSigma, \nu , \bm{c} \} $ and $p_j(\gamma)\vcentcolon=\Pr \left(  
	\lvert \htheta_j \rvert < c - t_{\gamma,\nu}  \hsigma_{j} 
	\right)$.
	From \eqref{eq:level:tost}, we know that $p(\alpha)\leq\alpha$. For a solution to exist, it suffices to show the conditions under which $\alpha < \alpha_{\max} \vcentcolon= \lim_{\gamma \to 0.5^-} p(\gamma)$, where $\gamma \to 0.5^-$ denotes the limit to $0.5$ from below. 
	We first find a lower bound to $p_j(\gamma)$ when $\gamma\to0.5^-$. 
	We have 
	\begin{equation*}
		\begin{aligned}
			\lim_{\gamma \to 0.5^-} p_j(\gamma) &= \lim_{\gamma \to 0.5^-}Q_{\nu}\left(-t_{1-\gamma,\nu},\, \frac{\lambda_j(\gamma) - c}{\sigma_{j}},\, \frac{c \sqrt{\nu} }{\sigma_{j} t_{1-\gamma,\nu}} \right) - Q_{\nu}\left(t_{1-\gamma,\nu},\, \frac{\lambda_j(\gamma) + c}{\sigma_{j}},\, \frac{c \sqrt{\nu} }{\sigma_{j} t_{1-\gamma,\nu}} \right)\\ 
			&= \Phi\left(-\frac{\lambda_j(\gamma)-c}{\sigma_j}\right)-\Phi\left(-\frac{\lambda_j(\gamma)+c}{\sigma_j}\right)
			\geq 
			\Phi\left(\frac{2c}{\sigma_{\max}}\right)-0.5,
		\end{aligned}
	\end{equation*}
	where $Q_{\nu}(t,\, y,\, z)$ corresponds to a special case of Owen's Q-functions, $\sigma_{\max} = \max_{j\in\{1,\dots,m\} }\sigma_j$, and the last inequality is obtained by replacing $\lambda_j(\gamma)\in[-c,c]$ by $c$ or $-c$. 
	
	Then, starting from \eqref{eq:joint:tost}, under independence we obtain
	\begin{equation*}
		\begin{aligned}
			\alpha_{\max}=\lim_{\gamma \to 0.5^-} p(\gamma) &= \lim_{\gamma \to 0.5^-} \Pr \left(  
			\bigcap_{j=1}^m \left\{ -\frac{\lambda_j(\gamma)+c}{\hsigma_j} + t_{\gamma,\nu}\leq T_{\nu,j}\leq -\frac{\lambda_j(\gamma)-c}{\hsigma_j} - t_{\gamma,\nu} \right\}
			\right) \\
			&= \lim_{\gamma \to 0.5^-} \prod_{j=1}^m \Pr \left(  
			-\frac{\lambda_j(\gamma)+c}{\hsigma_j} + t_{\gamma,\nu}\leq T_{\nu,j}\leq -\frac{\lambda_j(\gamma)-c}{\hsigma_j} - t_{\gamma,\nu}
			\right)\\
			&= \prod_{j=1}^m \lim_{\gamma \to 0.5^-}  p_j(\gamma)
			\geq 
			\left[\Phi\left(\frac{2c}{\sigma_{\max}}\right)-1/2\right]^m .
		\end{aligned}
	\end{equation*}
	Thus, $\alpha<\alpha_{\max}$ is satisfied if $\alpha<\left[\Phi\left(\frac{2c}{\sigma_{\max}}\right)-1/2\right]^m$. This in turn implies
	\begin{equation*}
		\alpha^{1/m}+1/2 < \Phi\left(\frac{2c}{\sigma_{\max}}\right) < 1 \quad \Longleftrightarrow \quad \sigma_{\max} < \frac{2c}{\Phi^{-1}\left(\alpha^{1/m}+1/2\right)} \quad \text{with} \quad  \alpha < (1/2)^m.
	\end{equation*}
	When $\sigma_{\max}$ is unknown (i.e., when $\bSigma$ is unknown), it is simply replaced by its estimate $\widehat{\sigma}_{\max}$ in the above condition to study the existence of $\widehat{\alpha}^*$.
	
	\section{Asymptotic Results for the Multivariate $\alpha$-TOST}
	\label{appendix:asymptotic_alphaTOST}
	
	We aim to show that 
	\begin{equation}
		\widehat{\alpha}^*=\alpha^*+o_p\left(\nu^{-1}\right).\label{rate:alpha}
	\end{equation}
	In order to do so, we need to use the following results
	\begin{align}
		\widehat{\btheta} &=\btheta + \mathcal{O}_p\left(\nu^{-1/2}\right), \label{rate:theta}\\
		\widehat{\sigma}_{j} &= {\sigma}_{j} + \mathcal{O}_p\left(\nu^{-1}\right),\label{rate:sigma}\\
		\widehat{\Sigma}_{i,j} &= \Sigma_{i,j} + \mathcal{O}_p\left(\nu^{-3/2}\right). \label{rate:Sigma}
	\end{align}
	The results in \eqref{rate:theta} and \eqref{rate:sigma} directly follow from Equations (B.1) and (B.2) of Boulaguiem et al.,\cite{boulaguiem23} where the result holds element-wise assuming $\lim_{\nu\to\infty}\sqrt{\nu}\sigma_{j}=v_j$ where $v_j$ is constant for all $j=1,\dots,m$. Equation~\eqref{rate:Sigma} stems from the central limit theorem. Indeed, defining $\bGamma=\nu\bSigma$, we have $\sqrt{\nu}\left(\widehat{\Gamma}_{i,j} -\Gamma_{i,j}\right) \xrightarrow{d}\mathcal{N}\left(0,V_{i,j}\right)$ for all $i,j \in \{1,\dots,m\}$, where ${V}_{i,j}$ is constant and $\xrightarrow{d}$ denotes convergence in distribution. This implies that $\widehat{\Gamma}_{i,j} = \Gamma_{i,j}+\mathcal{O}_p\left(\nu^{-1/2}\right)$, which in turn yields the result in \eqref{rate:Sigma}. Next, we define $\btau \vcentcolon= ( \sigma_1,\dots,\sigma_m,\Sigma_{1,2},\dots,\Sigma_{1,m},\Sigma_{2,3},\dots,\Sigma_{m-1,m} )^T$. The covariance matrices $\bSigma$ and $\hbSigma$ can be viewed as functions of $\btau$ and $\widehat{\btau}$ respectively, and we can redefine the size function as $p \{ \alpha, \blambda(\alpha), \btau, \nu , \bm{c} \} \vcentcolon=p \{ \alpha, \blambda(\alpha), \bSigma, \nu , \bm{c} \} $ and $p \{ \alpha, \blambda(\alpha), \widehat{\btau}, \nu , \bm{c} \} \vcentcolon= p \{ \alpha, \blambda(\alpha), \hbSigma, \nu , \bm{c} \} $. Using this notation, we can see that equations \eqref{rate:sigma} and \eqref{rate:Sigma} give us the convergence in probability of $\widehat{\bm{\tau}}$ to $\bm{\tau}$, and in particular that 
	\begin{equation}
		\max_h\left|\widehat{\tau}_h - \tau_h\right|=\mathcal{O}_p(\nu^{-1}),
		\label{rate:tau}
	\end{equation}
	for any $h=1,\dots,H$ where $H=m(m-1)/2$.
	
	Additionally, standard regularity conditions need to be assumed, namely the sequences 
	\begin{equation*}
		\left\{\frac{\partial}{\partial \alpha} \,  p(\alpha, \blambda(\alpha), \btau, \nu , \bm{c} ) \right\}_{\nu \in \mathcal{M}}\;\;\; \text{and} \;\;\; \left\{\frac{\partial}{\partial \tau_h} \,  p(\alpha, \blambda(\alpha), \btau, \nu , \bm{c} ) \right\}_{\nu \in \mathcal{M}},
	\end{equation*}
	converge uniformly in $\alpha$ and $\tau_h$, where $\mathcal{M} \subseteq \real$.
	Using \eqref{rate:tau} and the continuity in $\bm{\tau}$ of $p(\alpha, \blambda(\alpha), \btau, \nu , \bm{c} )$, by the continuous mapping theorem we have that $p(\alpha,\blambda(\alpha), \widehat{\bm{\tau}}, \nu, \bm{c})=p( \alpha,\blambda(\alpha), \bm{\tau}, \nu, \bm{c})+o_p(1)$, which by Lemma~5.10 of van der Vaart\cite{van2000asymptotic} yields
	\begin{equation}
		\widehat{\alpha}^* =\alpha^*+o_p\left(1\right).
		\label{consist:alpha}
	\end{equation}
	Next, for a given point $[ \widetilde{\alpha},\widetilde{\bm{\tau}}^T ] ^T$ between $[ \widehat{\alpha}^*,\widehat{\bm{\tau}}^T ] ^T $ and $[ \alpha^*,\bm{\tau}^T ] ^T$, an application of the mean value theorem provides
	\begin{equation*}
		p(\widehat{\alpha}^*,\blambda(\widehat{\alpha}^*),\widehat{\bm{\tau}},\nu,\bm{c}) -  p\left({\alpha}^*,\blambda({\alpha}^*),\bm{\tau},\nu,\bm{c}\right) = \frac{\partial}{\partial \alpha} p\left(\widetilde{\alpha},\blambda(\widetilde{\alpha}),\widetilde{\btau},\nu,\bm{c}\right) \left(\widehat{\alpha}^* - {\alpha}^*\right) + \sum_{h=1}^{H}\frac{\partial}{\partial \tau_h} p\left(\widetilde{\alpha},\blambda(\widetilde{\alpha}),\widetilde{\bm{\tau}},\nu,\bm{c}\right) \left(\widehat{\tau}_h - \tau_h\right).
	\end{equation*}
	where 
	\begin{equation*}
		\frac{\partial}{\partial \alpha} p\left(\widetilde{\alpha},\blambda(\widetilde{\alpha}),\widetilde{\bm{\tau}},\nu,\bm{c}\right)
		= \left.
		\frac{\partial}{\partial \alpha} p\left(x,\blambda(x),\widetilde{\bm{\tau}},\nu,\bm{c}\right)\right\vert_{x = \widetilde{\alpha}}\, \;\;\;  \; \text{and} \;\;\;\;\;  
		\frac{\partial}{\partial \tau_h} p\left(\widetilde{\alpha},\blambda(\widetilde{\alpha}),\widetilde{\bm{\tau}},\nu,\bm{c}\right)
		= \left.
		\frac{\partial}{\partial \tau_h} p\left(\widetilde{\alpha},\blambda(\widetilde{\alpha}),y,\nu,\bm{c}\right)\right\vert_{y = \widetilde{\tau}_h}.
	\end{equation*}
	This implies 
	\begin{equation}
		\frac{\partial}{\partial \alpha} p\left(\widetilde{\alpha},\blambda(\widetilde{\alpha}),\widetilde{\bm{\tau}},\nu,\bm{c}\right) \left(\widehat{\alpha}^* - {\alpha}^*\right) + \sum_{h=1}^{H}\frac{\partial}{\partial \tau_h} p\left(\widetilde{\alpha},\blambda(\widetilde{\alpha}),\widetilde{\bm{\tau}},\nu,\bm{c}\right) \left(\widehat{\tau}_h - \tau_h\right) = 0,
		\label{eq:atost:meanval}
	\end{equation}
	since from \eqref{eq:alpha_TOST_pop}, we have $p(\widehat{\alpha}^*,\blambda(\widehat{\alpha}^*),\widehat{\bm{\tau}},\nu,\bm{c}) =  p\left({\alpha}^*,\blambda(\alpha^*),\bm{\tau},\nu,\bm{c}\right) = \alpha$.
	
	By the uniform convergence of $\frac{\partial}{\partial \alpha} p\left( \alpha,\blambda(\alpha), {\bm{\tau}}, \nu, \bm{c}\right)$ and $\frac{\partial}{\partial \tau_h} p\left( \alpha,\blambda(\alpha), {\bm{\tau}}, \nu, \bm{c}\right)$, we use the second Theorem~7.17 of Rudin\cite{rudin1953principles} to get
	\begin{equation*}
		\lim_{\nu \to \infty} \frac{\partial}{\partial \alpha} \; p\left( \widetilde{\alpha},\blambda(\widetilde{\alpha}), {\widetilde{\bm{\tau}}}, \nu, \bm{c}\right) = \frac{\partial}{\partial \alpha} \; \lim_{\nu \to \infty} p\left( \widetilde{\alpha},\blambda(\widetilde{\alpha}), {\widetilde{\bm{\tau}}}, \nu, \bm{c}\right) = \frac{\partial}{\partial \alpha} \alpha= 1 ,
	\end{equation*}
	and 
	\begin{equation*}
		\lim_{\nu \to \infty} \frac{\partial}{\partial \tau_h} p\left( \widetilde{\alpha},\blambda(\widetilde{\alpha}), {\widetilde{\bm{\tau}}}, \nu, \bm{c}\right)  = \frac{\partial}{\partial \tau_h} \lim_{\nu \to \infty} p\left( \widetilde{\alpha},\blambda(\widetilde{\alpha}), {\widetilde{\bm{\tau}}}, \nu, \bm{c}\right) = \frac{\partial}{\partial \tau_h} \alpha= 0,
	\end{equation*}
	which in turn yields
	\begin{equation}
		\lim_{\nu \to \infty}\frac{\frac{\partial}{\partial \tau_h} p\left(\widetilde{\alpha},\blambda(\widetilde{\alpha}),\widetilde{\bm{\tau}},\nu,\bm{c}\right)}{\frac{\partial}{\partial \alpha} p\left(\widetilde{\alpha},\blambda(\widetilde{\alpha}),\widetilde{\bm{\tau}},\nu,\bm{c}\right)} = 0,
	\end{equation}
	for all $h=1,\dots,H$.
	
	Thus, rearranging the terms in \eqref{eq:atost:meanval} and for sufficiently large $\nu$ we get
	\begin{align*}
		\left|\widehat{\alpha}^* - {\alpha}^*\right| &= 
		\left| \sum_{h=1}^{H}\frac{\frac{\partial}{\partial \tau_h} p\left(\widetilde{\alpha},\blambda(\widetilde{\alpha}),\widetilde{
				\btau},\nu,\bm{c}\right)}{\frac{\partial}{\partial \alpha} p\left(\widetilde{\alpha},\blambda(\widetilde{\alpha}),\widetilde{\btau},\nu,\bm{c}\right)}\left(\widehat{\tau}_h - \tau_h\right) \right| \\
		&\leq \sum_{h=1}^{H}\left|\frac{\frac{\partial}{\partial \tau_h} p\left(\widetilde{\alpha},\blambda(\widetilde{\alpha}),\widetilde{\btau},\nu,\bm{c}\right)}{\frac{\partial}{\partial \alpha} p\left(\widetilde{\alpha},\blambda(\widetilde{\alpha}),\widetilde{\btau},\nu,\bm{c}\right)} \right| \left|\left(\widehat{\tau}_h - \tau_h\right)\right| \\
		&= o_p(\max_h\left|\widehat{\tau}_h - \tau_h\right|)=o_p(\nu^{-1}),
	\end{align*}
	where the last equality stems from $\eqref{rate:tau}$.
	Therefore, this result verifies $\eqref{rate:alpha}$ and concludes the proof.
	
	\section{Convergence Rate and Uniqueness of the Multivariate $\alpha$-TOST}
	\label{appendix:uniqueness_alphaTOST}
	Algorithm~\ref{app:alg:multiv_alphaTOST} describes the way we solve the multivariate $\alpha$-TOST in \eqref{eq:alpha_TOST_pop}.
	
	\begin{algorithm}[H] \label{app:alg:multiv_alphaTOST}
		\SetAlgoLined
		\Input{nominal significance level $\alpha$, covariance matrix $\bOmega$, degrees of freedom $\nu$, equivalence margins $\bm{c}$, maximum number of iterations $r_{\max}$, algorithmic tolerance $\epsilon_{min}$; }
		\Output{adjusted significance level $\alpha^*$; }
		$\alpha^{*(0)} = \alpha$\;
		$\blambda (\alpha^{*(0)}) = \argsup_{\btheta \in \Theta_0 } p( \alpha^{*(0)}, \btheta, \bOmega, \nu, \bm{c}  )$\;
		$r = 0$\;
		\While{
			$ \lvert p\left\{\alpha^{*(r)}, \blambda(\alpha^{*(r)}), \bOmega, \nu , \bm{c} \right\} -\alpha \rvert > \epsilon_{\min}$ and $r \leq r_{\max}$ }{
			$\alpha^{(r)}_{0} = \alpha^{*(r)}$\;
			$ \alpha^{(r)}_1 = \alpha^{(r)}_{0} + \alpha - p\left\{\alpha^{(r)}_{0}, \blambda(\alpha_0^{(r)}), \bOmega, \nu , \bm{c} \right\}$\;
			$k = 1$\;
			\While{ $ \lvert \alpha^{(r)}_k - \alpha^{(r)}_{k-1} \rvert > \epsilon_{\min}$ }
			{$k = k+1$\;
				$ \alpha^{(r)}_k = \alpha^{(r)}_{k-1} + \alpha - p\left\{\alpha^{(r)}_{k-1}, \blambda(\alpha_0^{(r)}), \bOmega, \nu , \bm{c} \right\}$\;
				
			}
			$r = r+1$\;
			$\alpha^{*(r)} = \alpha^{(r-1)}_{k}$\;
			$\blambda (\alpha^{*(r)}) = \argsup_{\btheta \in \Theta_0 } p \left\{ \alpha^{*(r)}, \btheta, \bOmega, \nu, \bm{c}  \right\}$;
		}        
		\caption{Multivariate $\alpha$-TOST}
	\end{algorithm}
	We aim to show that the inner loop converges to a unique solution exponentially fast
	\begin{equation*}
		\Big| \alpha^{(r)}_k  - \alpha^{*(r)}  \Big| < \frac{1}{2} \exp(-bk).
	\end{equation*}
	Fixing $\alpha$, $\bSigma$, $\nu$, $\bm{c}$ and $\blambda(\alpha_0^{(r)})$ for an arbitrary $r\in\mathbb{N}$, we simplify the notation and define  
	$p(\gamma)\vcentcolon=p\left\{\gamma,\blambda(\alpha_0^{(r)}),\bSigma,\nu,\bm{c}\right\}$.
	We restrict our attention to a subspace of the solution space $\mathcal{A}\vcentcolon= \left\{x \in [\alpha, \, 0.5) \,\big| \; p\left(x\right) > 0\right\}$, and we assume that $p\left(\gamma\right)$ is continuously differentiable such that $0 < \dot{p}\left(\gamma\right) < 2$, where 
	\begin{equation*}
		\dot{p}\left(\gamma\right) \vcentcolon= \left. \frac{\partial}{\partial x} p\left(x\right) \right\vert_{x = \gamma}\,. 
	\end{equation*}
	
	Let us now define
	\begin{equation}
		\label{eq:banach:func}
		T(\gamma) \vcentcolon= \alpha + \gamma - p\left(\gamma\right),
	\end{equation}
	and obtain as a result of the mean value theorem, for any $\alpha_1,\alpha_2\in\mathcal{A}$,
	\begin{equation*}
		\big|T(\alpha_1) - T(\alpha_2)\big| = \big|1 - \dot{p}\left(\widetilde{\alpha}\right)\big|\,  \big|\alpha_1 - \alpha_2\big| < \big|\alpha_1 - \alpha_2\big|,
	\end{equation*}
	where $\widetilde{\alpha} = \delta \alpha_1 + (1-\delta) \alpha_2$ for some $\delta \in [0,1]$.
	We can then use Kirszbraun theorem\cite{federer2014geometric} to extend the function $T(\alpha)$ to a contraction map from $\real$ to $\real$ with respect to $\alpha\in\mathcal{A}$, and Banach fixed point theorem will ensure that the sequence $\{\alpha_k\}$ defined by $\alpha_k=T(\alpha_{k-1})$ for $k\geq 1$, will converge to the unique fixed point $\alpha^*$, i.e.,  $T(\alpha^*)=\alpha^*$.
	Thus, by evaluating \eqref{eq:banach:func} at $\alpha^*$ and rearranging the terms we get 
	\begin{equation*}
		\alpha^* =
		\argzero_{\gamma \in \mathcal{A}} \; p\left(\gamma\right) - \alpha =
		\argzero_{\gamma \in [\alpha, 0.5)} \; p\left(\gamma\right) - \alpha.
	\end{equation*}
	This of course requires that a solution exists in $\mathcal{A}$ which is seldom unmet in practical cases where $\sigma_{\max}$ is not too large or $\nu$ not too small. 
	As a result, there exists some $0<\epsilon<1$ such that for $k\geq 1$ we have 
	\begin{equation*}
		\Big|\alpha_k - \alpha^*\Big| < \epsilon^k \big|\alpha^* - \alpha\big| < \frac{1}{2} \exp(-bk). 
	\end{equation*}
	
	\section{Additional Simulation Results}
	\label{appendix:simulations} 
	
\begin{figure*}[ht]
		\centering
		\includegraphics[width=0.8\textwidth]{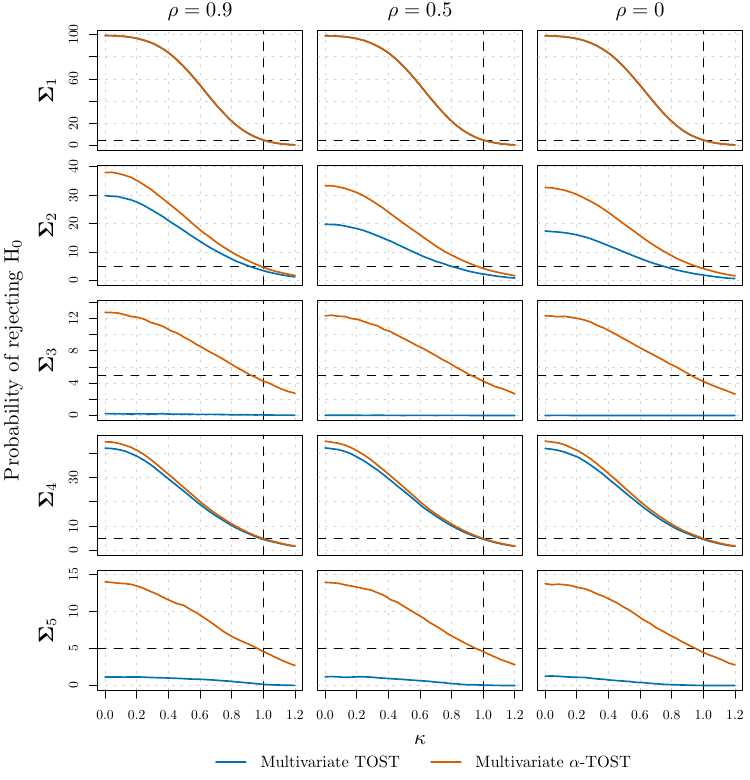}
		\caption{Probability of rejecting the null hypothesis (y-axis) as a function of $\kappa \in [0, 1.2]$ (x-axis) for different sets of variances (rows) and levels of dependence (columns) for the conventional multivariate TOST (blue line) and the multivariate $\alpha$-TOST (orange line) in the settings considered in Simulation 2 (refer to the second column of Table~\ref{fig:sim_table} for details).}
		\label{fig:sim2}
	\end{figure*}

\begin{figure*}[ht]
		\centering
		\includegraphics[width=0.8\textwidth]{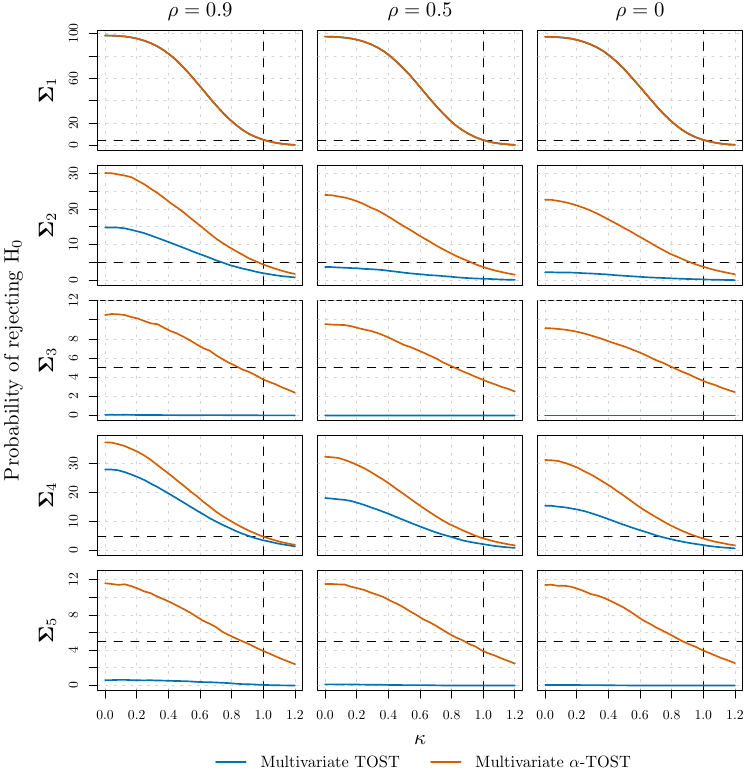}
		\caption{Probability of rejecting the null hypothesis (y-axis) as a function of $\kappa \in [0, 1.2]$ (x-axis) for different sets of variances (rows) and levels of dependence (columns) for the conventional multivariate TOST (blue line) and the multivariate $\alpha$-TOST (orange line) in the settings considered in Simulation 3 (refer to the third column of Table~\ref{fig:sim_table} for details).}
		\label{fig:sim3}
	\end{figure*}

\begin{figure*}[ht]
		\centering
		\includegraphics[width=0.8\textwidth]{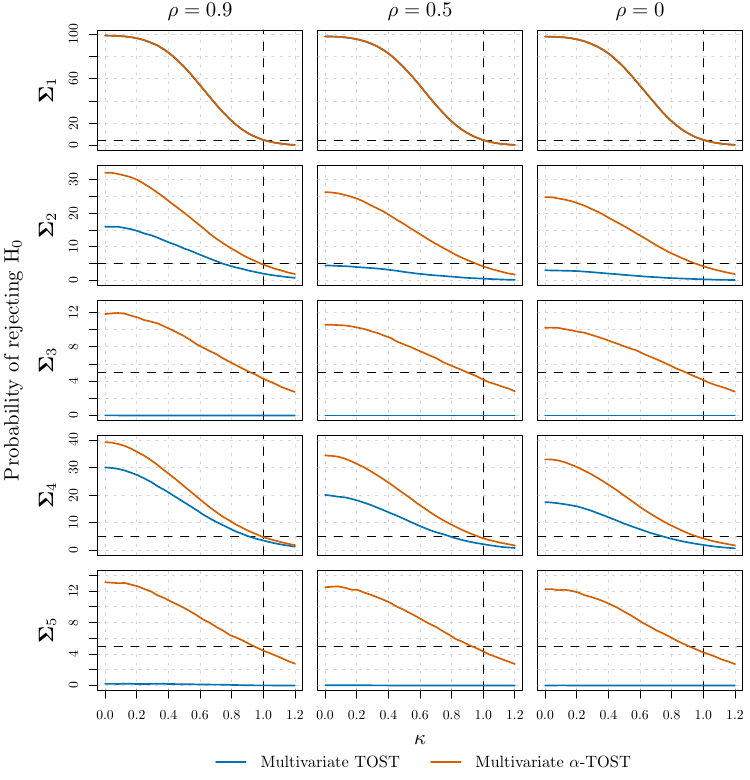}
		\caption{Probability of rejecting the null hypothesis (y-axis) as a function of $\kappa \in [0, 1.2]$ (x-axis) for different sets of variances (rows) and levels of dependence (columns) for the conventional multivariate TOST (blue line) and the multivariate $\alpha$-TOST (orange line) in the settings considered in Simulation 4 (refer to the fourth column of Table~\ref{fig:sim_table} for details).}
		\label{fig:sim4}
	\end{figure*}

	In this section, we present the results of three additional simulation settings described in Table~\ref{fig:sim_table}.
	The results are presented in Figures~\ref{fig:sim2} to \ref{fig:sim4}.
	
	\section{Details on the Application Study}
	\label{app:application}    
	
\begin{figure}[h]
		\centering
		\includegraphics[width=0.8\textwidth]{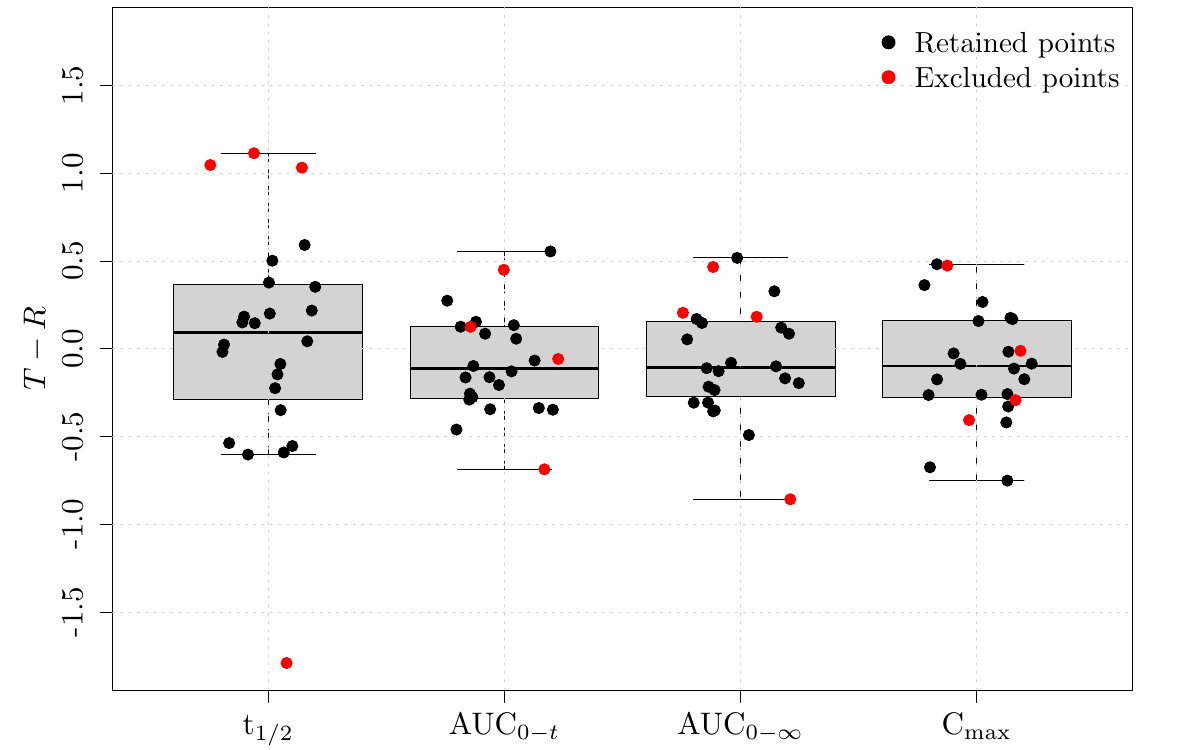}
		\caption{Boxplots for each pharmacokinetic measurement of interest, where red points represent the ones excluded from the analysis.}
		\label{fig:app_box}
	\end{figure}

\begin{figure}
		\centering
		\includegraphics[width=0.9\textwidth]{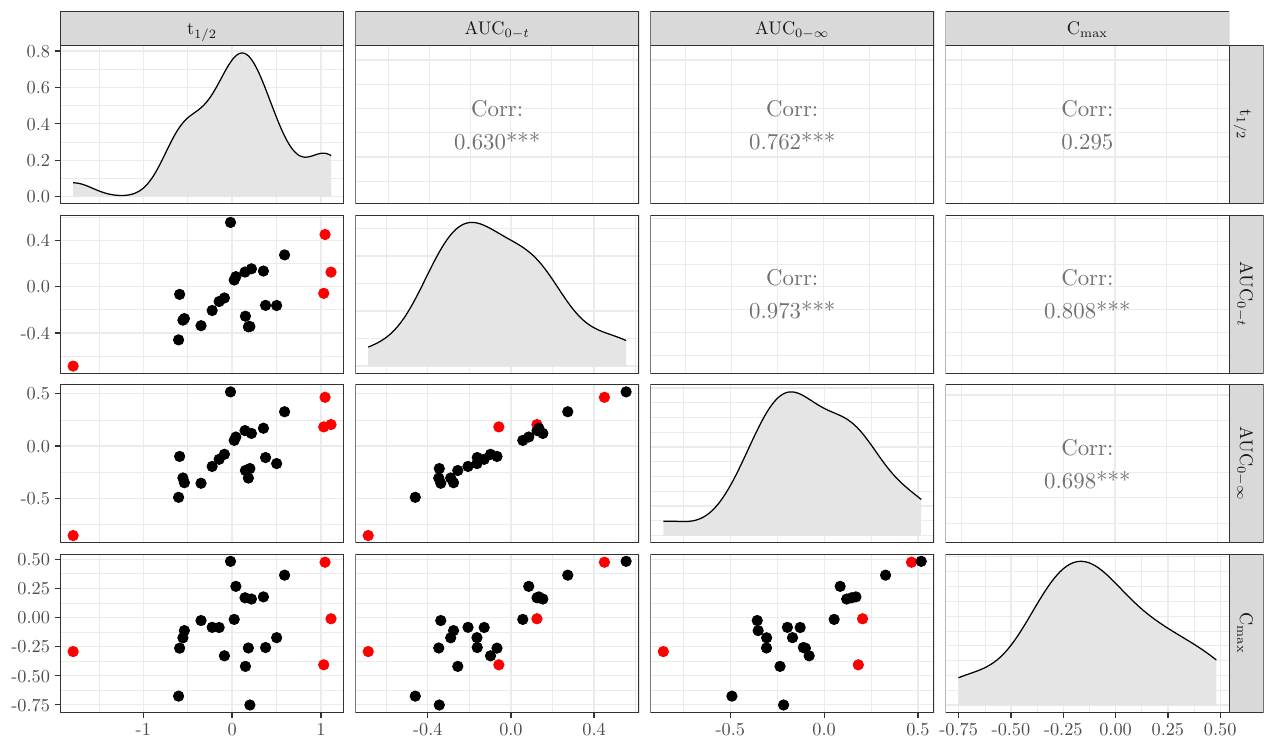}
		\caption{Scatterplots, kernel density estimates and correlations of the (logarithmically transformed) differences between the two formulations of ticlopidine hydrochloride for each pair of the pharmacokinetic measurements. The symbols $^{***}$, $^{**}$ and $^{*}$ indicate $p$-values smaller than 0.001, 0.01 and 0.05, respectively, for testing H$_0: \rho_{i,j} = 0$ vs. H$_1: \rho_{i,j} \neq 0$, where $\rho_{i,j}$ corresponds to the correlation between the variables in row $i$ and column $j$. The figure was obtained using the default parameters of the \texttt{ggpairs} function from the \texttt{GGally} package in \texttt{R}.\cite{ggpairs} Red points represent the ones excluded from the analysis.}
		\label{fig:pairscatter_all}
	\end{figure}

	Figure~\ref{fig:app_box} shows boxplots for the considered variables before outlier removal (i.e.,~$n=24$).
	This highlights the potential presence of outlying cases for the variable t$_{1/2}$. To mitigate this issue, one could employ a robust (multivariate) estimator of location and scatter to identify outlying cases and remove them from the analysis. 
	Nevertheless, this is not feasible due to the small $n/m$ ratio.
	We thus applied a univariate screen for the presence of outliers in t$_{1/2}$, and excluded these points from the analysis.
	Namely, 
	we first computed a robust $z_i$-score for such variable, say $z_i = \{ x_i - \operatorname{median}(x) \} / \operatorname{mad}(x)$, where $\operatorname{mad}(\cdot)$ represents the median absolute deviation and it is adjusted by a consistency correction factor (which is $\approx$1.4826 for normal data).
	Then, we retained only those points whose  $ -z_{\alpha/2} \leq z_i \leq z_{\alpha/2} $, where $z_{\alpha/2}$ denotes the upper $\alpha/2$ quantile of a standard normal distribution (for $\alpha=0.05$).
	This led to the removal of 4 points from the analysis, which are highlighted in red in Figures~\ref{fig:app_box} and \ref{fig:pairscatter_all}.
	
	\vskip 2cm
	
\end{document}